\documentclass[10pt,journal]{IEEEtran} 
\usepackage{comment}
\usepackage[linesnumbered,ruled,vlined]{algorithm2e}
\usepackage{mathtools, cases}
\usepackage{textcomp}
\usepackage{makecell}
\includecomment{toexclude}
\usepackage{amsthm, enumitem}
\usepackage{amsmath,amsfonts,amssymb}
\usepackage{cite, url}
\usepackage{verbatim}
\usepackage{bm}
\usepackage{graphicx}
\usepackage{xcolor}
\usepackage{subcaption}
\usepackage{epstopdf}
\usepackage{mathrsfs}
\usepackage{txfonts}
\usepackage{lineno}
\usepackage{multirow}
\usepackage{booktabs}
\usepackage{color}
\usepackage{multicol}
\usepackage{stfloats}
\usepackage{diagbox}
\usepackage{esint}
\usepackage{float}
\usepackage{placeins}
\usepackage{algpseudocode}
\usepackage{empheq}
\usepackage{tabularx}

\allowdisplaybreaks[4]

\usepackage{xcolor}

\usepackage{geometry}
\begin{document}

\title{LGVSC: A Large-Model-Driven Generative Video Semantic Communication Framework}

\author{Yu Ma, Hang Yin, Li Qiao, Shuo Sun, 
Zhen Gao,~\IEEEmembership{Senior Member,~IEEE}, Yin Xu,~\IEEEmembership{Senior Member,~IEEE}, and Wenjun Zhang,~\IEEEmembership{Fellow,~IEEE}
 \vspace{-5mm}
\thanks{Copyright (c) 2026 IEEE. Personal use of this material is permitted. However, permission to use this material for any other purposes must be obtained from the IEEE by sending a request to pubs-permissions@ieee.org.}
\thanks{Yu Ma is with the School of Information and Electronics, Beijing Institute of Technology, Beijing 100081, China, also with Zhongguancun Academy, Beijing 100094, China (e-mail: yu.ma@bit.edu.cn).}
\thanks{Hang Yin, Shuo Sun are with the School of Information and Electronics, Beijing Institute of Technology, Beijing 100081, China (e-mail: \{yh, sunshuo2002\}@bit.edu.cn).}
\thanks{Li Qiao is with the Department of Electrical and Computer Engineering, The University of Hong Kong, Pokfulam Road, Hong Kong (e-mail: qiaoli@hku.hk).}
\thanks{Zhen Gao is with Beijing Institute of Technology (BIT), Zhuhai 519088, China, also with the State Key Laboratory of CNS/ATM, Beijing 100081, China, also with the MIIT Key Laboratory of Complex-Field Intelligent Sensing, Beijing 100081, China, also with the Advanced Technology Research Institute, BIT, Jinan 250307, China, also with the Yangtze Delta Region Academy, BIT, Jiaxing 314019, China (e-mail: gaozhen16@bit.edu.cn).}
\thanks{Yin Xu and Wenjun Zhang are with the School of Information Science and Electronic Engineering, Shanghai Jiao Tong University, Shanghai 200240, China (e-mail: \{xuyin, zhangwenjun\}@sjtu.edu.cn).}
}

\maketitle
\begin{abstract}
Driven by the massive video transmission requirements in the Internet of Everything, semantic communication holds great promise for striking a balance between transmission efficiency and quality.
This paper introduces a large-model-driven generative video semantic communication (LGVSC) framework, enabling efficient video semantic transmission under extremely low bandwidth conditions. 
First, by decoupling the encoder and decoder as well as exposing explicit intermediate semantic representations, LGVSC maintains interpretability, avoiding the black-box behavior commonly observed in end-to-end systems.
Next, we introduce a new metric, i.e., the probability-based semantic similarity score (PSSS), which quantifies semantic similarity for complex modalities within a continuous range, allowing for more precise evaluation of semantic content.
Building on PSSS, we propose a semantic-guided keyframe extraction module driven by a multimodal large model. This module can enhance fine-grained semantic consistency during keyframe selection at the transmitter, optimizing transmission bandwidth without compromising semantic fidelity.
Additionally, we design a generative large-model-driven dynamic semantic-adaptive decoder at the receiver, which can adapt to videos of arbitrary lengths.
Simulation results demonstrate that LGVSC significantly outperforms traditional schemes, {achieving a channel bandwidth ratio on the order of $\bm{10^{-4}}$ to $\bm{10^{-3}}$,} while maintaining strong zero-shot generalization across downstream tasks.
\end{abstract}

\IEEEpeerreviewmaketitle

\section{Introduction}

\IEEEPARstart{S}{emantic} {communication} (SemCom) has emerged as a promising paradigm for enhancing communication efficiency under bandwidth constraints, particularly as traditional Shannon-based syntactic communication approaches face limitations in supporting immersive communication. Recent studies have focused on two main directions: (1) enhancing compression efficiency by transmitting only essential semantic information, thereby enabling ultra-low bitrate communication; and (2) integrating artificial intelligence to facilitate task-oriented transmission and enhancing semantic accuracy. In particular, the paradigm evolution from specialist models to large models has been shown to substantially benefit semantic-aware communication system design~\cite{ying2026specialist}. SemCom demonstrates strong potential under conditions of extremely limited bandwidth and low signal-to-noise ratio (SNR), such as concurrent high-resolution video streaming in the Internet of Everything (IoE).

Despite these advancements, two fundamental challenges remain unresolved. \textbf{First}, the efficient semantic extraction and transmission for arbitrary-length videos remains an open challenge. Specifically, generative video semantic communication (GVSC)~\cite{yinYinhangGenerativeVideoSemantic2025} can only transmit short video clips ($8-16$ frames) at $10\textsuperscript{-2}$ channel bandwidth ratio (CBR\footnote{$\mathrm{CBR} = \frac{B}{C \times H \times W \times F}$, 
where $B$ denotes transmitted symbols and $C$, $H$, $W$, and $F$ denote the numbers of color channels, height, width, and frame, respectively.}). Wireless deep video semantic transmission (DVST)~\cite{wangDVSTWirelessDeepVideo2023} offers higher fidelity through entropy coding and deep joint source-channel coding (JSCC) but requires substantially higher CBR. Semantic video conferencing (SVC)~\cite{tongMultimodalSemanticCommunication2025} is designed exclusively for meeting-style video transmission and is thus unsuitable for general video applications.
\textbf{Second}, current SemCom frameworks exhibit limited generalization across different downstream tasks, often requiring costly task-specific fine-tuning to achieve effective performance~\cite{guoVideoQASCAdaptiveSemantic2025a}.
These limitations highlight the lack of scalable, bandwidth-efficient, and training-free SemCom frameworks for long video sequences.

\begin{figure*}[t]
  \centering
  \begin{minipage}{0.27\textwidth}
    \scriptsize %
    \centering
    
    \begin{subfigure}{0.48\linewidth}
      \includegraphics[width=\linewidth]{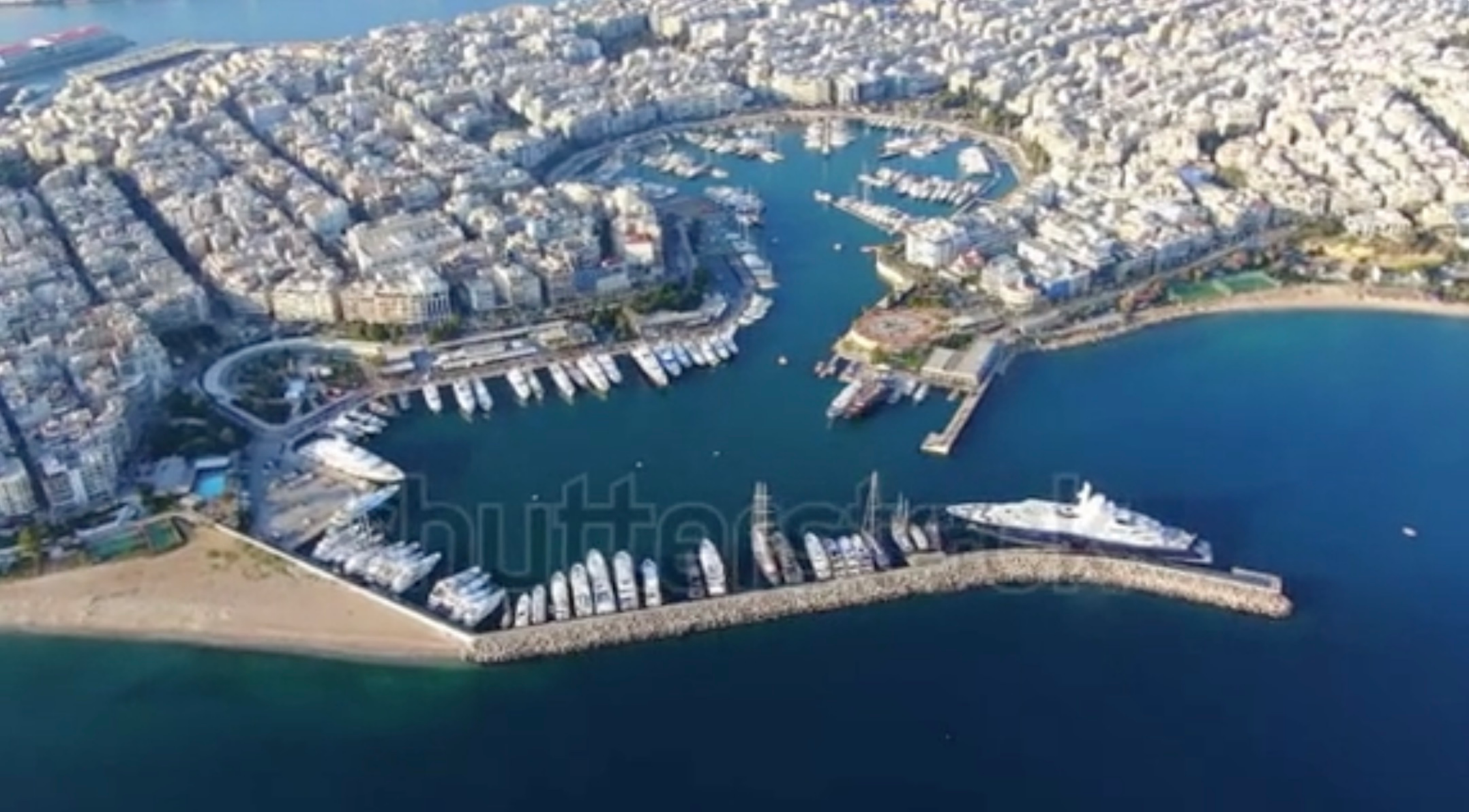}
      \caption{Origin}
    \end{subfigure}
    \hfill
    \begin{subfigure}{0.48\linewidth}
      \includegraphics[width=\linewidth]{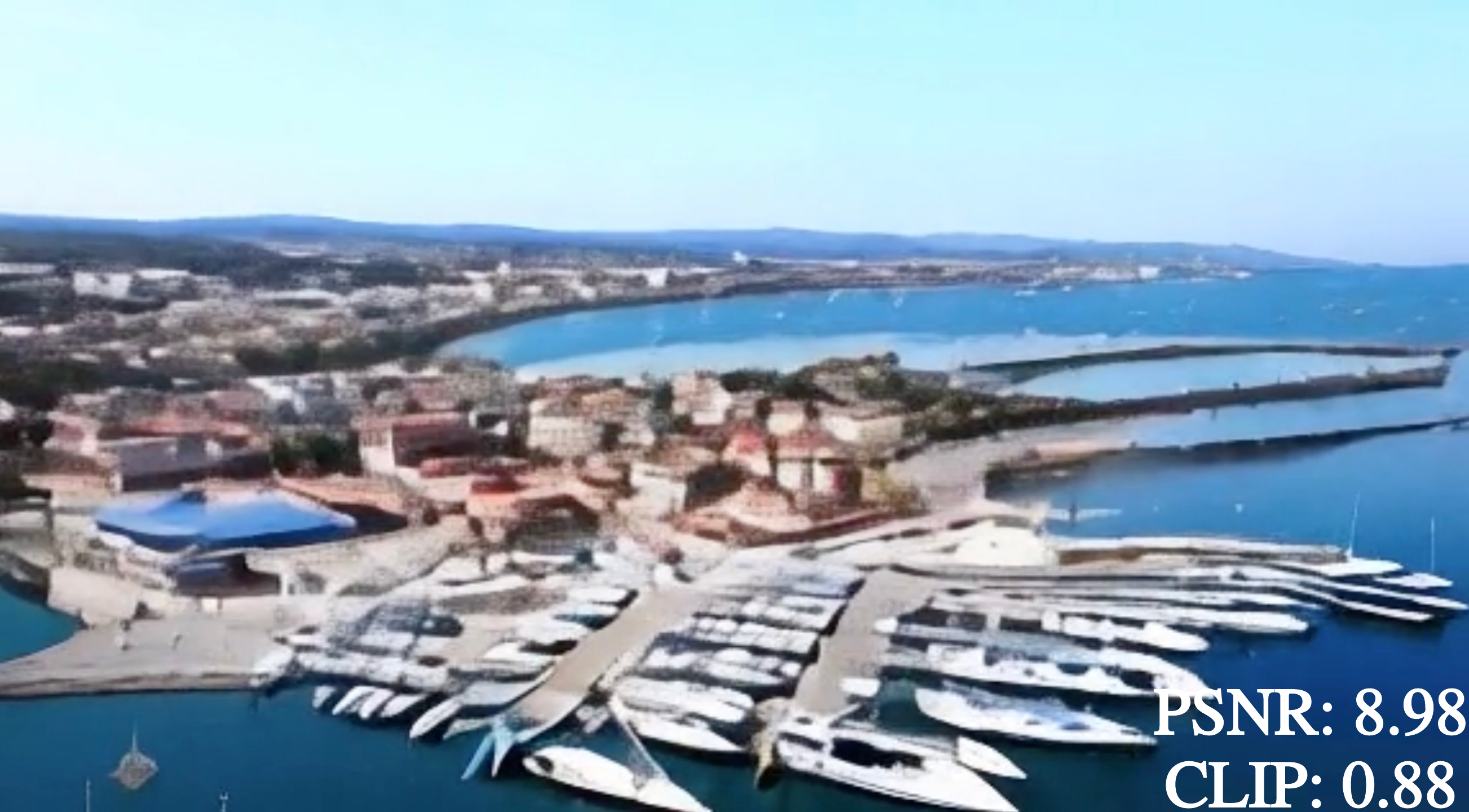}
      \caption{Text-Only}
      \label{subfig:desc}
    \end{subfigure}

    \begin{subfigure}{0.48\linewidth}
      \includegraphics[width=\linewidth]{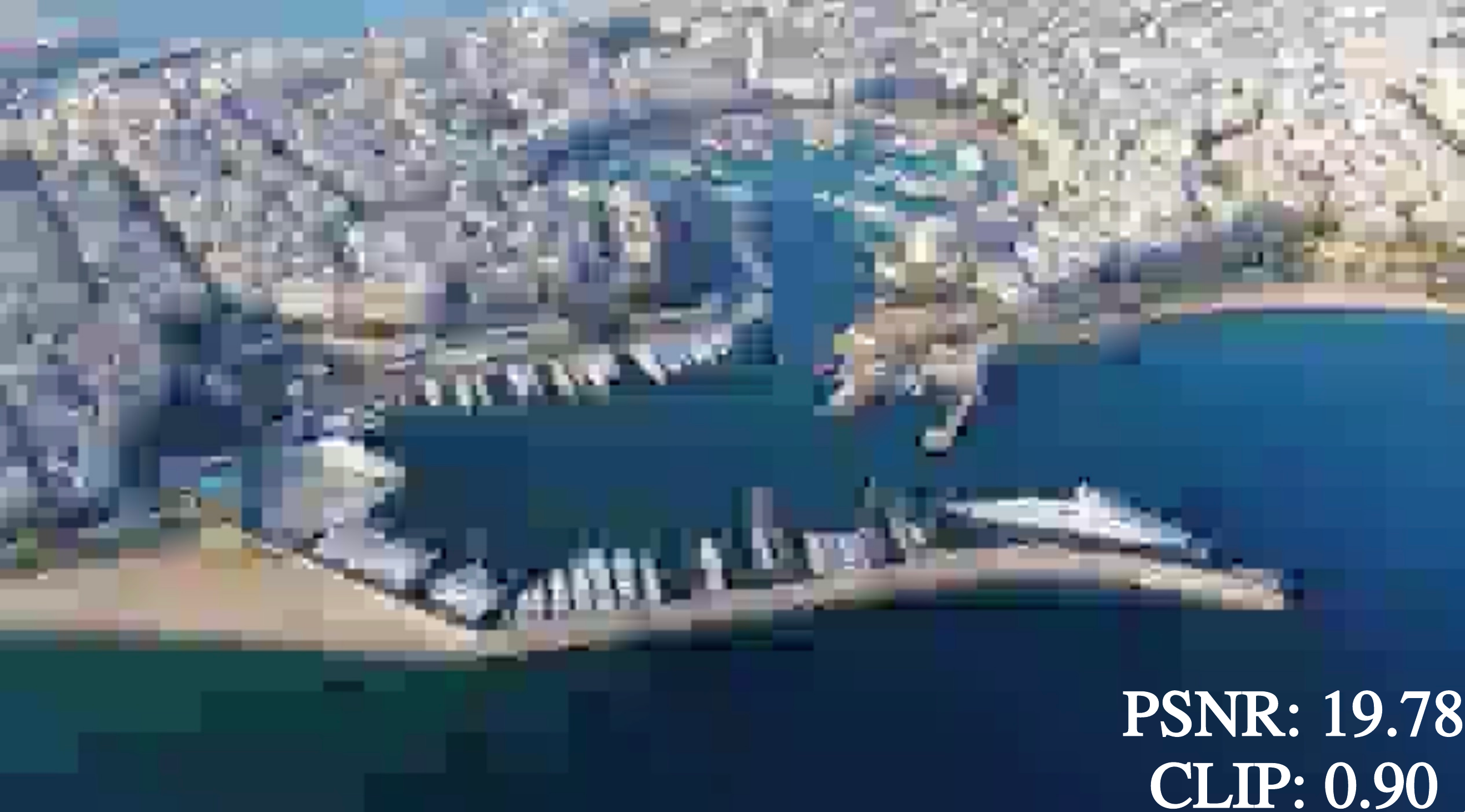}
      \caption{H.264}
      \label{subfig:264}
    \end{subfigure}
    \hfill
    \begin{subfigure}{0.48\linewidth}
      \includegraphics[width=\linewidth]{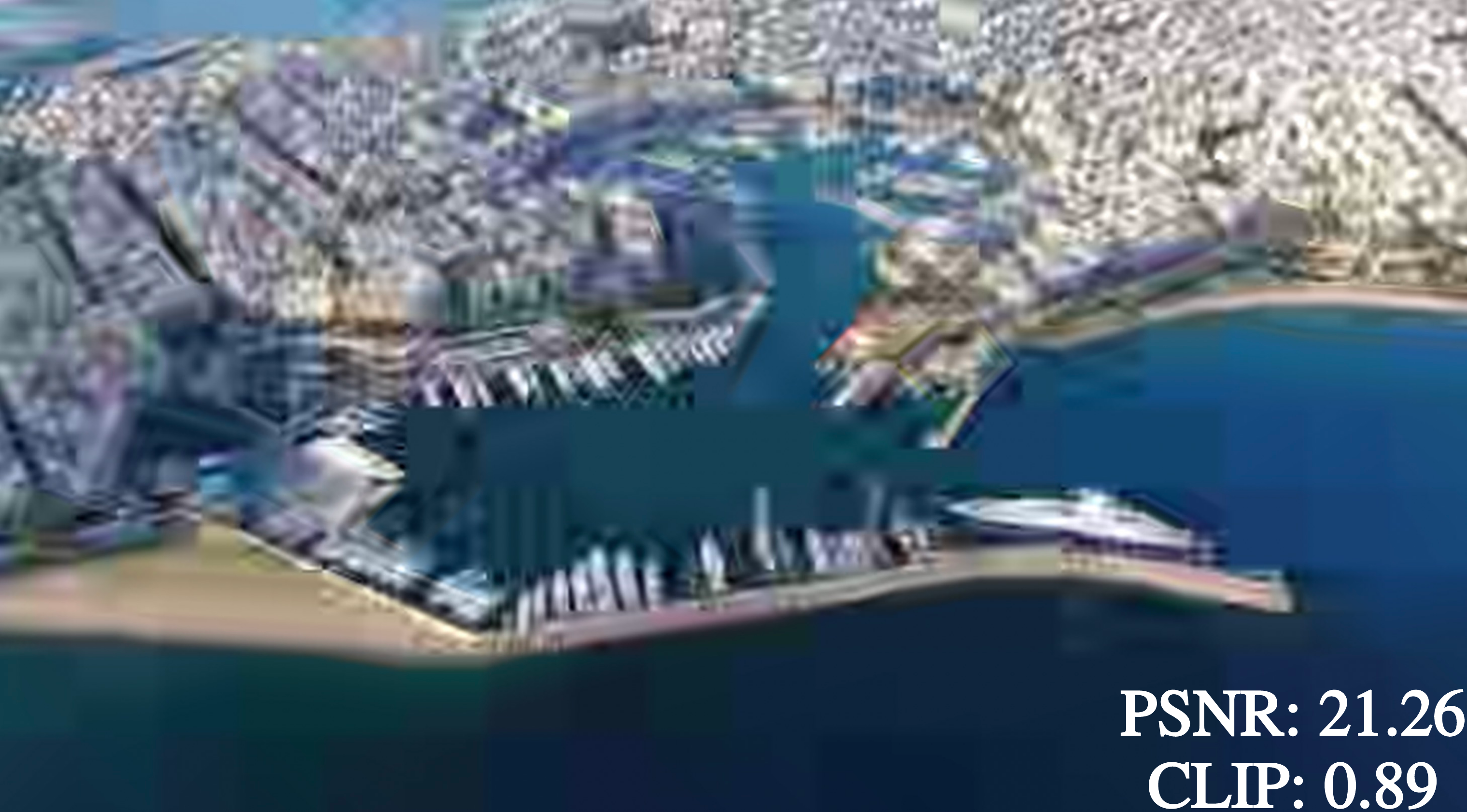}
      \caption{H.265}
      \label{subfig:265}
    \end{subfigure}

    \begin{subfigure}{0.48\linewidth}
      \includegraphics[width=\linewidth]{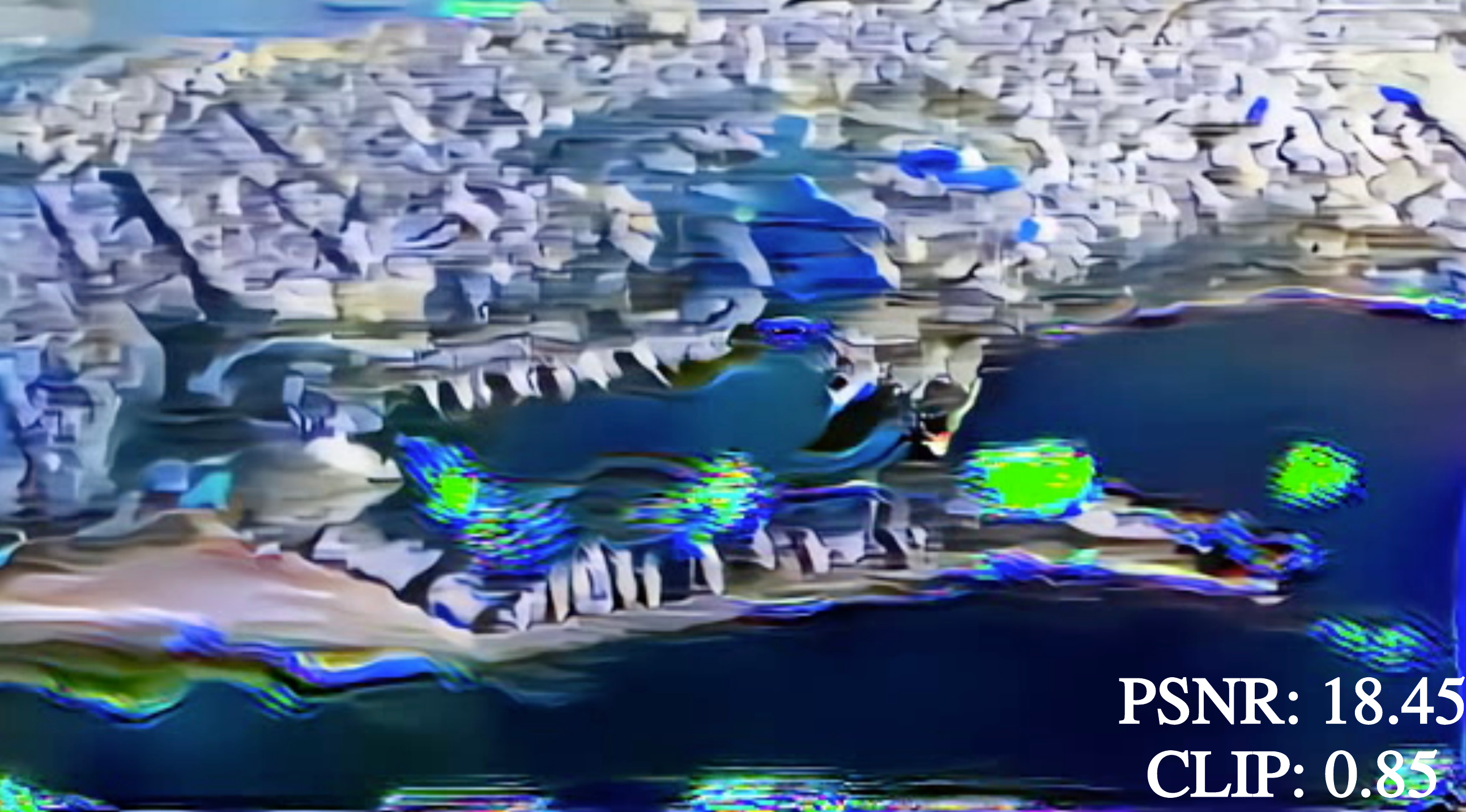}
      \caption{DVST}
      \label{subfig:dvst}
    \end{subfigure}
    \hfill
    \begin{subfigure}{0.48\linewidth}
      \includegraphics[width=\linewidth]{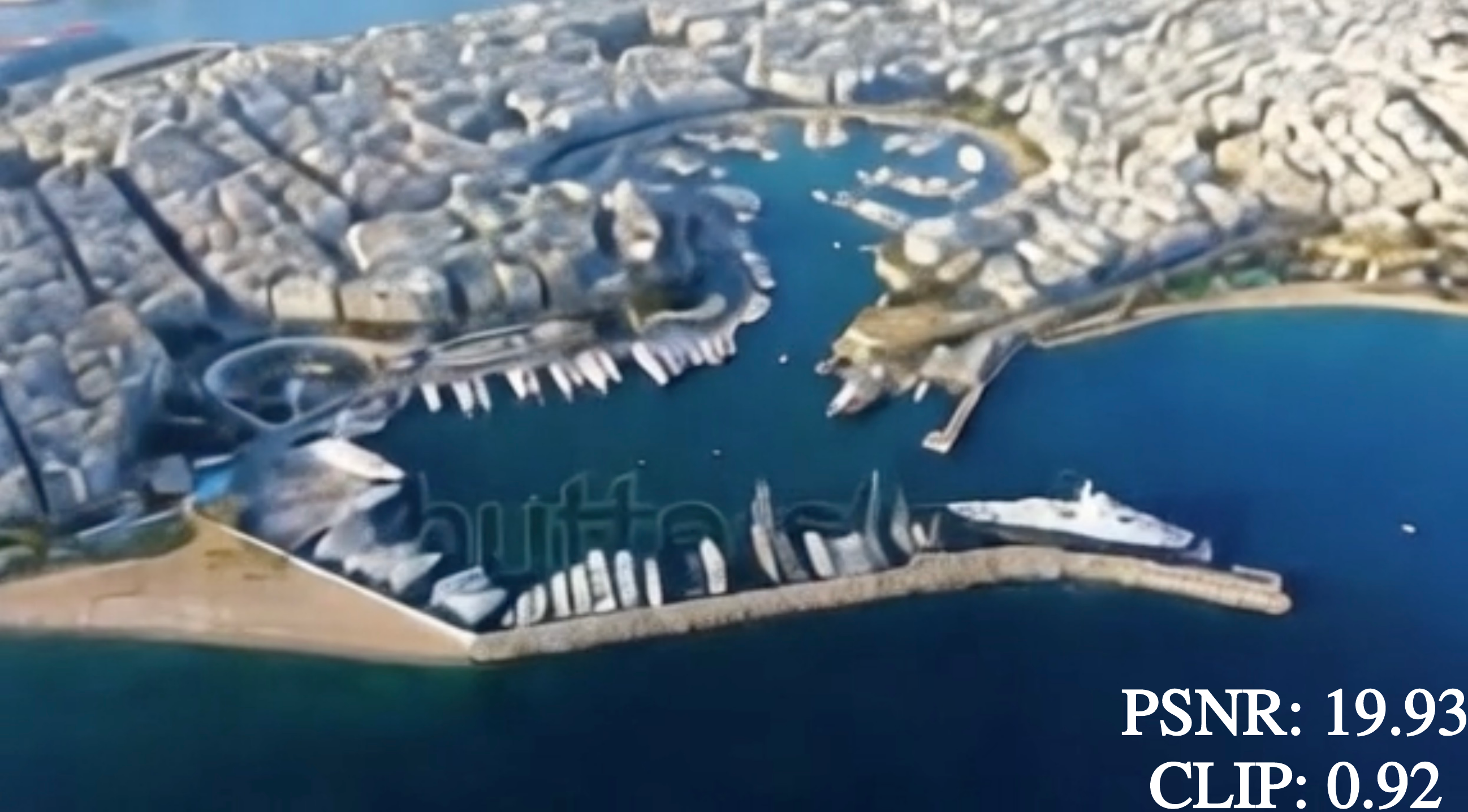}
      \caption{LGVSC(Ours)}
      \label{subfig:ours}
    \end{subfigure}
    \captionsetup{font={footnotesize}, singlelinecheck=off, justification=raggedright, name={Fig.}, labelsep=period}
    \caption{Visual comparison of reconstructed video frames under different transmission schemes at SNR=10 dB.}
    \label{fig:visual_comparison}
  \end{minipage}
  \hfill
  \begin{minipage}{0.72\textwidth}
    \centering
    \includegraphics[width=0.9\linewidth]{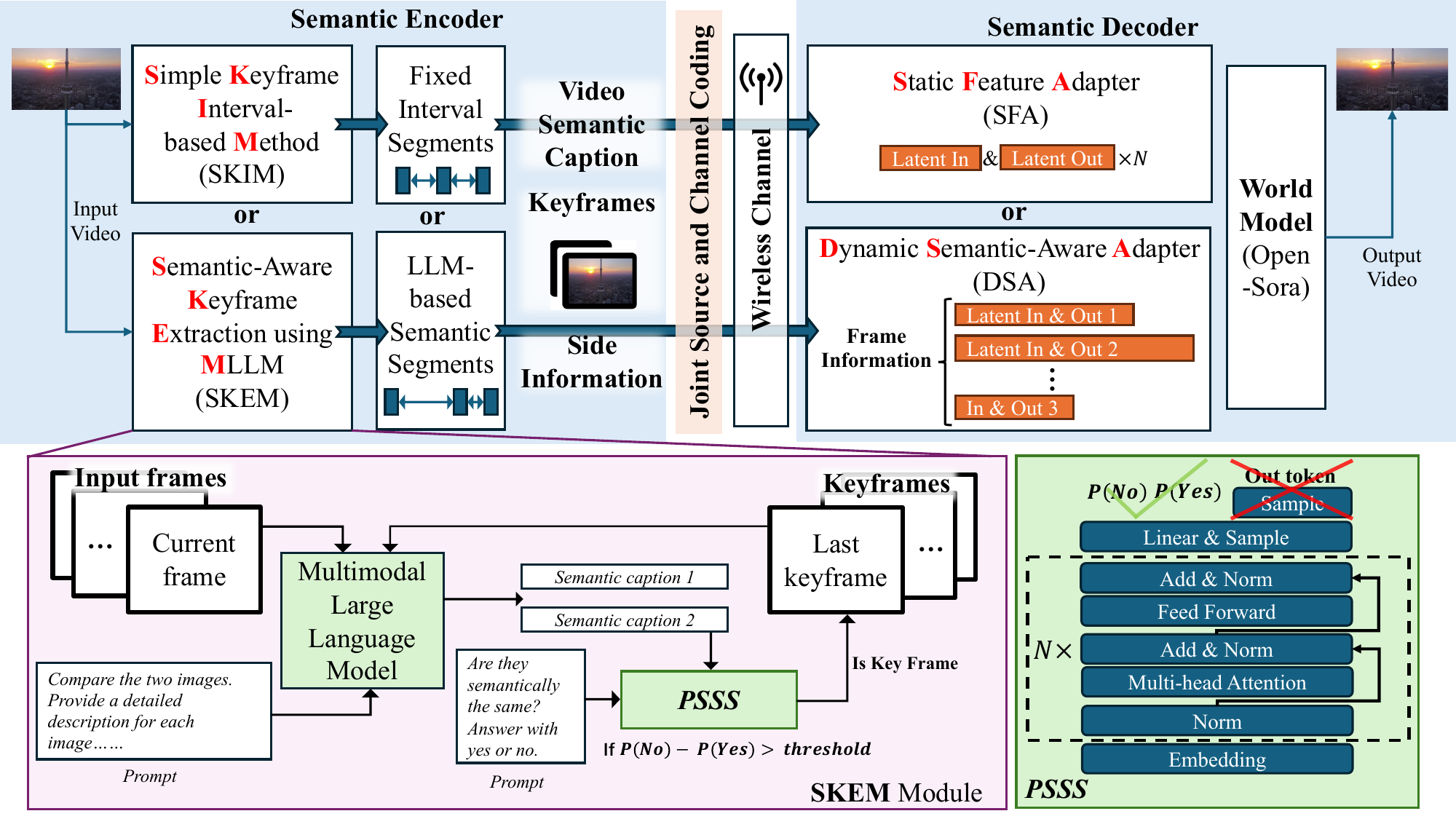}
    \captionsetup{font={footnotesize}, singlelinecheck=off, justification=raggedright, name={Fig.}, labelsep=period}
    \caption{Proposed LGVSC framework using SKEM (or SKIM) at transmitter and DSA (or SFA) at receiver.}
    \label{fig:SystemDiag}
  \end{minipage}
  
\end{figure*}

Recent advances in large models have demonstrated strong zero-shot capabilities across modalities~\cite{qiaoTokenCommunicationsLarge2025}. Moreover, the integration of large models into 6G networks has been shown to enable embodied intelligence for integrated perception, communication, and computation~\cite{li2025large}. We observe that large models are capable of achieving state-of-the-art performance in video SemCom without fine-tuning. A representative case is that world models like Sora~\cite{liu2024sorareviewbackgroundtechnology} act as effective semantic decoders, capable of reconstructing realistic video from text-only semantic information under low CBR (Fig.~\ref{subfig:desc})~\cite{yinYinhangGenerativeVideoSemantic2025}~\cite{qiaoLatencyAwareGenerativeSemantic2024}, in contrast to traditional methods that exhibit noticeable distortion (Fig.~\ref{subfig:265},~\ref{subfig:264}, and~\ref{subfig:dvst}).

While video reconstruction or generation based solely on textual semantics can preserve high-level meaning, they often fail to capture fine-grained and long-term semantic dependencies essential for complex video content. For images, neural semantic coding schemes such as nonlinear transform source-channel coding (NTSCC)~\cite{NTSCC} have shown strong capability in preserving fine-grained visual semantics under extremely low CBR. However, extending image-level semantic coding to long video sequences incurs prohibitive communication overhead. Enabling efficient video semantic transmission therefore requires selective delivery of semantically critical multimodal information, which is fundamentally constrained by the lack of a principled and quantifiable semantic similarity metric for adaptive temporal information selection.

Therefore, we propose a large-model-driven generative video semantic communication (LGVSC) framework, which introduces a novel semantic similarity evaluation method supported by a multimodal large model. The proposed LGVSC can improve semantic transmission quality, aligning with human perception, while {reducing the CBR for video SemCom to the order of $10^{-4}$ to $10^{-3}$}. The framework designs large-model-driven multimodal semantic encoders and decoders, decoupling the semantic perception process from traditional communication pipelines, thereby preserving semantic interpretability and enhancing compatibility with existing wireless networks. A core design goal of LGVSC is to support the transmission and reconstruction of \emph{arbitrary length} videos, which is a capability that existing GVSC systems, restricted to fixed-length clips, do not provide. This goal motivates all three proposed contributions and distinguishes LGVSC from prior art.
Our contributions are summarized as follows:

\begin{itemize}
  \item We propose a probability-based semantic similarity score (PSSS). This method accurately evaluates multimodal semantic similarity via predicted probability distributions, providing a continuous similarity score, thus offering more precise semantic comparison.

  \item Building on PSSS, we develop a multimodal large-model-driven semantic-aware keyframe extraction module (SKEM) at the transmitter. This module can explicitly incorporate semantic similarity into the autoregressive keyframe selection process, enhancing fine-grained semantic consistency in LGVSC.
  
  \item We design a diffusion world model-driven dynamic semantic-adaptive decoder. We conceive a dynamic semantic-aware adapter (DSA), {which serves as the architectural enabler for arbitrary length video reconstruction by loading latent variables of varying dimensions matched to each semantic segment.}
\end{itemize}

\section{Proposed LGVSC Framework}\label{sec:ourwork}
This section elaborates on the proposed LGVSC, which comprises three main components: a semantic encoder, a semantic decoder, and a JSCC module, as illustrated in Fig.~\ref{fig:SystemDiag}.

\subsection{Procedure and Modules of LGVSC Framework}

At the transmitter, we conceive a multimodal large-model-driven semantic encoder and this encoder can segment the input video $\mathcal{X} \in \mathbb{R}^{F \times H \times W \times C}$ by extracting keyframes $\mathcal{X}[K_n,:,:,:]$ ($K_n \in \{1,2,\cdots, F\}$, $1 \leq n \leq N+1$, $N+1$ is the number of keyframes) and forming a set of variable-length semantic segments $\mathcal{X} = \{\mathcal{S}_1; \mathcal{S}_2; \cdots; \mathcal{S}_N\}$. Here $\{\cdot~;~\cdot\}$ denotes concatenation along the temporal dimension, $\mathcal{S}_n = \mathcal{X}[K_n\!:\!K_{n+1}\!-\!1,:,:,:]$, and $N$ is the number of segments. Furthermore, each segment is encoded as $\{I^{\text{text}}_n, I^{\text{frame}}_n, I^{\text{side}}_n\}$: a semantic caption, the $n$-th keyframe (i.e., $I^{\text{frame}}_n = \mathcal{X}[K_n,:,:,:]$, here $K_{N+1} = F$ being the last frame), and a task-specific side information. $I^\text{text}_n$ is generated by a pre-trained model, providing compact spatiotemporal semantics. $\{I^\text{frame}_n\}_{n=1}^{N+1}$ can be extracted by simple keyframe interval-based method (SKIM) or SKEM, providing fine-grained semantics. PSSS, the core component of SKEM, will first be introduced in Section~\ref{subsec:psss}. Then, both SKIM and SKEM will be detailed respectively in Section~\ref{subsec:skim}.

For the semantic information from different modalities, namely $I^{\text{text}}_n$, $I^{\text{frame}}_n$, and $I^{\text{side}}_n$, we design distinct transmission strategies. The textual description and side information are encoded into binary bitstream. This stream is encoded by using a channel code and then modulated. The resulting signal is transmitted over the wireless channel. The receiver performs the corresponding reverse operations.
For each keyframe $I^{\text{frame}}_n$ ($1 \leq n \leq N$), we employ NTSCC for encoding. The frame undergoes nonlinear transformation and JSCC to produce the channel input vector $\mathbf{s}_n = f_{\text{NTSCC}}(I^{\text{frame}}_n)$, where $f_{\text{NTSCC}}$ denotes the NTSCC network function.

At the receiver side, the channel output vector $\hat{\mathbf{s}}_n$ is decoded to reconstruct the keyframe $\hat{I}^{\text{frame}}_n$. The semantic decoder reconstructs segments $\{\hat{\mathcal{S}}_n\}_{n=1}^N$ using a world model conditioned on $\{\hat{I}^{\text{text}}_n, \hat{I}^{\text{frame}}_n, \hat{I}^{\text{side}}_n\}$. A redesigned adapter enables alignment with our communication setting: the combination of SKIM and static feature adapter (SFA), i.e., SKIM-SFA, is suitable for low-latency, low-computational power and low dynamic variation scenarios, while the combination of SKEM and
dynamic semantic-aware adapter (DSA), i.e., SKEM-DSA, is ideal for scenarios requiring higher semantic fidelity in transmission. Both SFA and DSA will be detailed respectively in Section~\ref{subsec:dsa}. The final video is reconstructed by concatenating the restored video segments as
$\hat{\mathcal{X}} = \{\hat{\mathcal{S}}_1; \hat{\mathcal{S}}_2; \cdots; \hat{\mathcal{S}}_N\} \in \mathbb{R}^{F \times H \times W \times C}$.

\subsection{Proposed PSSS for Keyframe Selection}
\label{subsec:psss}

Transformer-based large models leverage their pre-trained knowledge to perform autoregressive reasoning in response to specific tasks. The attention mechanism allows these models to efficiently assess the relevance between the task and the input information.

To this end, we propose a semantic similarity evaluation prompt structured as follows: “$\langle$Info. A$\rangle$, $\langle$Info. B$\rangle$. Determine whether they are similar from the perspective of $\langle$Semantic Focus$\rangle$\footnote{In this context, “Info.” represents the extracted semantic information, while $\langle\text{Semantic Focus}\rangle$ defines the semantic aspect to be emphasized, e.g., concentrating on the subject or prioritizing the action.}, use yes or no to answer." As illustrated in Fig.~\ref{fig:SystemDiag}, in the PSSS module, existing approaches utilizing large models for evaluation typically rely on the sampled outputs of ``Yes'' or ``No'' as the final assessment results~\cite{TIFA}. However, such binary classification outputs cannot deliver a continuous measure of semantic similarity, limiting their ability to support fine-grained comparison across semantic segments.

To address this limitation, we skip the sampling step during inference. Instead, we use the model-predicted probability of the “Yes” token during autoregressive inference as a continuous-valued semantic similarity score $S_\mathrm{abs}$. This score is an absolute confidence value output by large models, i.e.,
\begin{equation}
\begin{split}
S_{\mathrm{abs}} = P\Bigl(
&\text{“Yes”} \mid \langle \text{Info. A} \rangle, \langle \text{Info. B} \rangle, \\
&\langle \text{Semantic Focus} \rangle
\Bigr) \in [0,1).
\end{split}
\end{equation}

However, $S_\mathrm{abs}$ is often influenced by imbalances in their training data. As a result, semantic similarity scores derived from this method may exhibit significant variation across different semantic contexts. To mitigate this issue, we further introduce a relative semantic similarity measure, expressed as
\begin{equation}
  S_\mathrm{rel}(\langle \text{Info. A} \rangle, \langle \text{Info. B} \rangle) = P(\text{``No"}) - P(\text{``Yes"}) \in (-1,1).
  \label{equa:2}
\end{equation}

This formulation normalizes the absolute similarity score by comparing the model's confidence in generating contradictory outputs for the same semantic input, effectively reducing the impact of semantic variability. The relative probability formulation in Eq.~(\ref{equa:2}) simultaneously reduces model pretraining bias and prompt formulation bias: any systematic shift in the model's overall response tendency affects both $P(\text{``Yes''})$ and $P(\text{``No''})$ proportionally, leaving $S_\mathrm{rel}$ largely unaffected. The same common-mode cancellation also yields graceful degradation under out-of-distribution inputs: when both frames are semantically degenerate, the model's confidence becomes diffuse and both token probabilities rise together, so $S_\mathrm{rel}\!\approx\!0$ and no spurious keyframe is inserted. Moreover, the $\langle$Semantic Focus$\rangle$ field provides a steerable, task-specific sensitivity: switching the focus between foreground action and background scene yields substantially different $S_\mathrm{rel}$ values for the same frame pair, enabling downstream-task-aware keyframe selection that a fixed visual-feature distance such as CLIP cannot provide. In section~\ref{sec:experiment}, this method is employed for the quantitative evaluation of semantic similarity.

\subsection{SKIM or SKEM-based Semantic Encoder at Transmitter}
\label{subsec:skim}
The semantic encoder consists primarily of a keyframe extraction module, segmenting the video signal into semantic segments. These segments are then encoded into semantic representations. To enable effective semantic encoding, we propose two keyframe extraction methods: SKIM and SKEM, which respectively adopt fixed-interval and semantic-guided segmentation strategies.

\subsubsection{SKIM}

To address low-latency and computationally constrained scenarios, SKIM is considered to segment the video into fixed time intervals. Given the number of semantic segments $N$, the video signal $\mathcal{X} \in \mathbb{R}^{F \times H \times W \times C}$ is divided into $N$ segments $\mathcal{X} = \{\mathcal{S}_1; \mathcal{S}_2; \cdots; \mathcal{S}_N\}$ of equal length $l_n = \frac{F}{N}$. 
Each segment's semantic text information $I^{\text{text}}_n$ is generated by multimodal large models, while the side information $I^{\text{side}}_n$ can be derived using optical flow~\cite{opticalflow}. The keyframe $I^{\text{frame}}_n$ is selected as the first frame of the segment. By adjusting $N$, we can control the number of transmitted keyframes to meet different CBR requirements.

The SKIM method is simple, with nearly real-time segmentation and minimal computational overhead. However, it accounts less for semantic content, potentially omitting critical semantic information and negatively impacting the reconstruction quality of the semantic decoder.
\subsubsection{SKEM}

To better align with the semantic information needs of the decoder's world model, we further propose SKEM. Based on a multimodal large language model, SKEM uses the PSSS method to evaluate the semantic similarity of video content, determining the segmentation of semantic segments.

The process is shown in the lower-left of Fig.~\ref{fig:SystemDiag}. SKEM operates in an autoregressive framework where the video signal $\mathcal{X} \in \mathbb{R}^{F \times H \times W \times C}$ is divided into a sequence of frames $\mathcal{X}[i,:,:,:]$, $1 \leq i \leq F$. The framework maintains a dynamic sequence of keyframes $\mathcal{X}[K_n,:,:,:]$, with the first keyframe $\mathcal{X}[K_1,:,:,:]$ automatically selected as $\mathcal{X}[1,:,:,:]$ with $K_1=1$. 

For the $n$-th ($2 \leq n \leq N$) keyframe selection with the corresponding autoregressive step, the current frame $\mathcal{X}\left[i,:,:,:\right] (K_{n-1}+1 \leq i \leq F)$ is compared to the latest keyframe $\mathcal{X}\left[K_{n-1},:,:,:\right]$ in the sequence, and a semantic similarity score $S_\mathrm{rel}(\mathcal{X}\left[i,:,:,:\right], \mathcal{X}\left[K_{n-1},:,:,:\right])$ according to~(\ref{equa:2}) is computed. Note that $S_\mathrm{rel}$ is a semantic \emph{similarity} score in the lower-is-more-similar sense: a lower value indicates that the two frames share greater semantic similarity (the model strongly prefers ``Yes''), while a higher value indicates greater semantic divergence. The threshold $\eta_{\text{th}}$ therefore functions as a \emph{semantic divergence threshold}: when $S_\mathrm{rel} > \eta_{\text{th}}$, the current frame is semantically sufficiently different from the latest keyframe to warrant insertion as a new keyframe; when $S_\mathrm{rel} \leq \eta_{\text{th}}$, the two frames are considered semantically similar and no new keyframe is inserted. This design is consistent with other lower-is-better distance metrics in the literature (e.g., LPIPS, AbsRel). If the similarity exceeds the threshold $\eta_{\text{th}}$, the current frame $\mathcal{X}\left[i,:,:,:\right]$ for $K_{n-1}+1 \leq i \leq F$ is added as the $n$-th keyframe and the search for the $\left(n+1\right)$-th keyframe begins, and so on. Otherwise, the process proceeds without modifying the keyframe sequence.

SKEM enhances frame selection by incorporating task-specific semantics, using pre-trained multimodal language models like InternVL~\cite{chen2024internvlscalingvisionfoundation} to generate textual descriptions $I^{\text{text}}_{i}~(K_{n-1} \leq i \leq F)$ and $I^{\text{text}}_{K_{n-1}}~(2 \leq n \leq N)$ for the current frame and latest keyframe. The PSSS method then computes the semantic similarity between these descriptions, simplifying the task by first converting the visual information into textual representation and then comparing them.

To get the frame semantics, we use the InternVL model, which can process multiple images simultaneously, ensuring that the model's description and focus remain aligned at each step, facilitating accurate semantic similarity evaluation.

Both the $\langle$Semantic Focus$\rangle$ and the semantic divergence threshold $\eta_{\text{th}}$ are dynamic inputs in the SKEM framework, allowing flexibility in adapting to various downstream tasks and video content. By adjusting $\eta_{\text{th}}$, the number of keyframes can be controlled to meet CBR requirements, ensuring efficient encoding. In practice, $\eta_{\text{th}}$ is calibrated against a target CBR budget on a small held-out set, analogously to quantization parameter selection in conventional video codecs, and no model retraining is required.
\subsection{SFA or DSA-based Semantic Decoder at Receiver}
\label{subsec:dsa}
The core of the semantic decoder lies the world model like Open-Sora~\cite{opensora}. We have redesigned the adapter component to enable the diffusion model within the world model to dynamically adapt to semantic segments $\mathcal{S}_n$ of variable lengths. The adapted model can controllably reconstruct the video segments based on adjacent keyframes and textual semantics.
\subsubsection{Static feature adapter (SFA)}
The SFA is an inherent component of the world model, where the segment length is fixed, and the latent feature dimension of the variational auto-encoder (VAE) encoder-decoder remains constant across all segments.

\subsubsection{Dynamic semantic adapter (DSA)}
To accommodate semantic segments of variable length, we further conceive the DSA, which dynamically adjusts the latent feature dimension of the VAE encoder-decoder based on the length of each semantic segment. When loading the world model, the model dimensions are correspondingly adapted, while the parameters remain reusable. 

\section{EXPERIMENT AND DISCUSSION}
\label{sec:experiment}
This section verifies the effectiveness of our proposed LGVSC framework that enables high-fidelity transmission of arbitrary length videos under ultra-low CBR. Particularly, we assess the zero-shot generalization of our LGVSC framework across three downstream tasks: video captioning, action classification, and depth estimation.

\subsection{Dataset}

We randomly selected 55 videos from the WebVid dataset~\cite{WebVid} as the test set in our experiments. The same set was used for both video captioning and depth estimation tasks. For action classification, we used the Kinetics-400 dataset~\cite{Kinetics400}, randomly selecting 14 videos for testing.

For the NTSCC model used to transmit keyframes, we trained it on one hundred thousand frames, using a learning rate of 0.0001, a batch size of 64, and 100 epochs. 
\subsection{Experimental settings}

We consider the following transmission schemes:

\textbf{SKEM+DSA}: Semantic segments are extracted using the SKEM method, with a semantic divergence threshold of $\eta_{\text{th}} = 0.35$. Keyframe information $I^\text{frame}_n$ is transmitted via NTSCC, textual description $I^{\text{text}}_n$ and side information $I^{\text{side}}_n$ are transmitted over an additive white Gaussian noise (AWGN) channel.

\textbf{SKIM+SFA}: Keyframes are extracted at fixed intervals, with $3-4$ frames randomly selected per video, matching the CBR of SKEM+DSA.

\textbf{Text-Only}: Only textual descriptions are transmitted.

\textbf{DVST}: Video transmission is done using the DVST method, with CBR set to its minimum. Specifically, Transmission parameters are $\eta = 0.2$ and $\lambda = 128$, where $\eta$ maps entropy to channel symbols and $\lambda$ controls the rate-distortion trade-off~\cite{wangDVSTWirelessDeepVideo2023}.

\textbf{H.264 or H.265 + LDPC}: H.264~\cite{h264} and H.265~\cite{h265} codecs are used for source encoding, while low-density parity-check (LDPC) is applied for channel encoding. Parameter details are listed in Table~\ref{tab:h26xldpc_params}. Quantization parameter (QP) controls the trade-off between video compression and quality, and group of pictures (GOP) size determines the distance between keyframes. Note that the cliff effect occurs when the SNR is below 6 dB.

The CBR values for each scheme at different SNR levels are summarized in Table~\ref{tab:snr_cbr}. The baseline methods exhibit lower compression efficiency (higher CBR) due to architectural limitations. {Note that the full LGVSC pipeline (SKEM+DSA and SKIM+SFA) operates around $6\times10^{-4}$, while the $10^{-4}$ value corresponds to the Text-Only configuration that transmits captions only.}

\begin{table}[tbp]
\caption{Parameter Settings for H.264/H.265 + LDPC
\label{tab:h26xldpc_params}}
\begin{center}
\scriptsize
\renewcommand{\arraystretch}{1.1}
\begin{tabular}{|>{\centering\arraybackslash}m{0.5cm}
                |>{\centering\arraybackslash}m{0.7cm}
                |>{\centering\arraybackslash}m{0.4cm}
                |>{\centering\arraybackslash}m{0.5cm}
                |>{\centering\arraybackslash}m{2.0cm}
                |>{\centering\arraybackslash}m{1.5cm}|}
\hline
\textbf{SNR (dB)} & \textbf{Codec} & \textbf{QP} & \textbf{GOP Size} & \textbf{LDPC Coding Rate} & \textbf{Modulation Order} \\
\hline
\multirow{2}{*}{6$^{\mathrm{*}}$}  & H.264 & 51 & 400 & 3/4 & 16-QAM  \\
\cline{2-6} %
   & H.265 & 51 & 400 & 3/4 & 4-QAM  \\
\hline
\multirow{2}{*}{8}  & H.264 & 51 & 80 & 1/2 & 16-QAM  \\
\cline{2-6}
   & H.265 & 51 & 50 & 1/2 & 16-QAM  \\
\hline
\multirow{2}{*}{10} & H.264 & 51 & 20  & 2/3 & 16-QAM \\
\cline{2-6}
   & H.265 & 50 & 13  & 2/3 & 16-QAM \\
\hline
\multicolumn{6}{l}{$^{\mathrm{*}}$The cliff effect occurs when the SNR is below 6 dB.}
\end{tabular}
\end{center}
\end{table}

\begin{table}[tbp]
\caption{CBR Values for Different SNR Levels
\label{tab:snr_cbr}}
\begin{center}
\scriptsize
\renewcommand{\arraystretch}{1.3}
\begin{tabular}{|>{\centering\arraybackslash}m{0.5cm} 
                |>{\centering\arraybackslash}m{1.0cm} 
                |>{\centering\arraybackslash}m{1.0cm} 
                |>{\centering\arraybackslash}m{0.8cm} 
                |>{\centering\arraybackslash}m{0.8cm} 
                |>{\centering\arraybackslash}m{1.8cm}|}
  \hline
  \textbf{SNR (dB)} & \textbf{SKIM + SFA} & \textbf{SKEM + DSA} & \textbf{Text-Only} & \textbf{DVST} & \textbf{H.264/H.265 + LDPC} \\
  \hline
  0  & $6.1\times{10}^{-4}$ & $6.1\times{10}^{-4}$ & ${10}^{-4}$ & --$^{\mathrm{*}}$ & -- \\
  \hline
  2  & $6.1\times{10}^{-4}$ & $6.1\times{10}^{-4}$ & ${10}^{-4}$ & -- & -- \\
  \hline
  4  & $6.3\times{10}^{-4}$ & $6.3\times{10}^{-4}$ & ${10}^{-4}$ & -- & -- \\
  \hline
  6  & $6.4\times{10}^{-4}$ & $6.4\times{10}^{-4}$ & ${10}^{-4}$ & $4\times{10}^{-3}$ & $1.70\times{10}^{-3}$\\
  \hline
  8  & $6.4\times{10}^{-4}$ & $6.4\times{10}^{-4}$ & ${10}^{-4}$ & $4\times{10}^{-3}$ & $1.40\times{10}^{-3}$ \\
  \hline
  10 & $6.6\times{10}^{-4}$ & $6.5\times{10}^{-4}$ & ${10}^{-4}$ & $4\times{10}^{-3}$ & $1.38\times{10}^{-3}$ \\
  \hline
  \multicolumn{6}{l}{$^{\mathrm{*}}$`--' indicates transmission failure or model collapse under low SNR conditions.}
\end{tabular}
\end{center}
\vspace{-5mm}
\end{table}

We tested the \textbf{SKEM+DSA} as our method in downstream evaluation. Both tests were conducted at SNR of 10 dB. For video captioning, we used the PLLava-7B~\cite{xu2024pllavaparameterfreellava} model; for action classification, we used TimeSformer~\cite{TimeSformer}; and for monocular depth estimation, we used Depth-Anything-v2~\cite{depth_anything_v2}. All models were configured according to their respective default settings.

\subsection{Evaluation metrics}

We evaluate the system using both traditional video transmission metrics and semantic-level metrics as follows.
The peak signal-to-noise ratio (PSNR) measures pixel-level image quality;
the structural similarity index measure (SSIM) assesses structural fidelity of images;
the contrastive language-image pre-training (CLIP) score measures semantic similarity between images, capturing high-level semantic features;
the learned perceptual image patch similarity (LPIPS) is a perceptual image quality metric based on human visual perception;
the deep image structure and texture similarity (DISTS) captures both structure and texture information using deep learning.
For each metric, the video quality is averaged across all frames. For the DVST method, PSNR and SSIM were tested on the average of the first 100 frames and the last 100 frames, demonstrating their cumulative error offset for long video transmission.

As for downstream tasks, we use the bilingual evaluation understudy (BLEU), the recall-oriented understudy for gisting evaluation longest common subsequence (ROUGE-L), and the bidirectional encoder representations from transformers (BERT) score~\cite{zhangbertscore} for video captioning; the Top-1 and Top-5 accuracy for action classification; the depth estimation metrics ($\text{D}_n$, the percentage of pixels where $\max\!\left( \frac{d_i}{d_i^*}, \frac{d_i^*}{d_i} \right) < 1.25^n$, for $n = 1, 2, 3$, $d_i^*$ is the ground truth), the absolute relative difference (AbsRel, $\frac{1}{|N|} \sum_{d \in N} \frac{|d - d^*|}{d^*}$, where $N$ is the number of available pixels in the ground-truth), and the root mean squared error (RMSE) for monocular depth estimation.
\subsection{Simulation results}

\begin{figure}[tbp]
    \centering

    \begin{subfigure}[t]{0.22\textwidth}
        \centering
        \includegraphics[width=\textwidth]{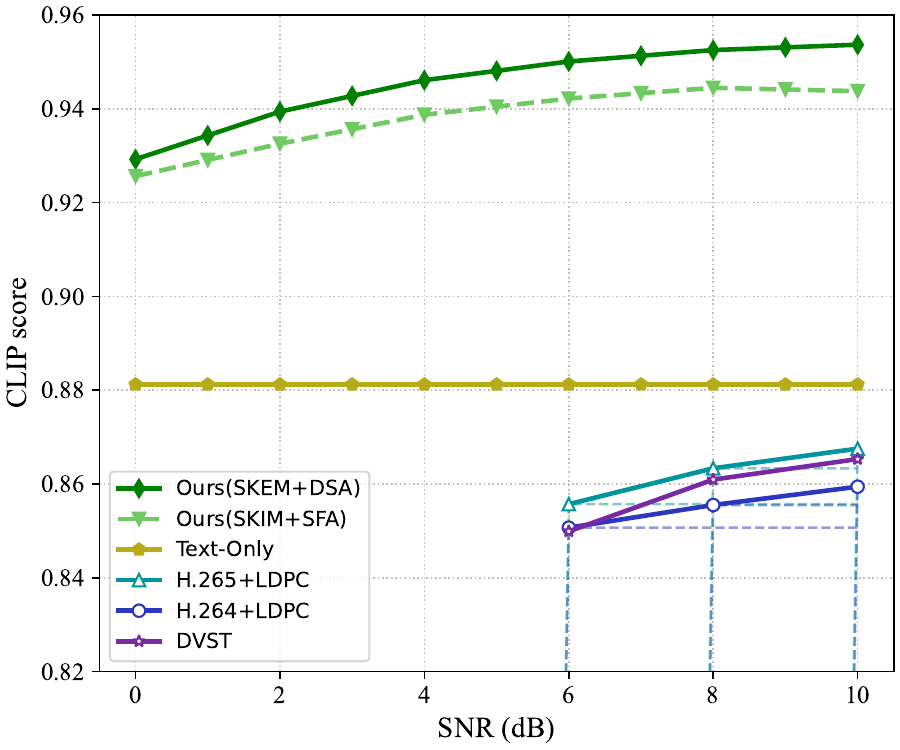}
        \caption{CLIP $\uparrow$}
        \label{fig:clip_vs_snr}
    \end{subfigure}

    \begin{subfigure}[t]{0.22\textwidth}
        \centering
        \includegraphics[width=\textwidth]{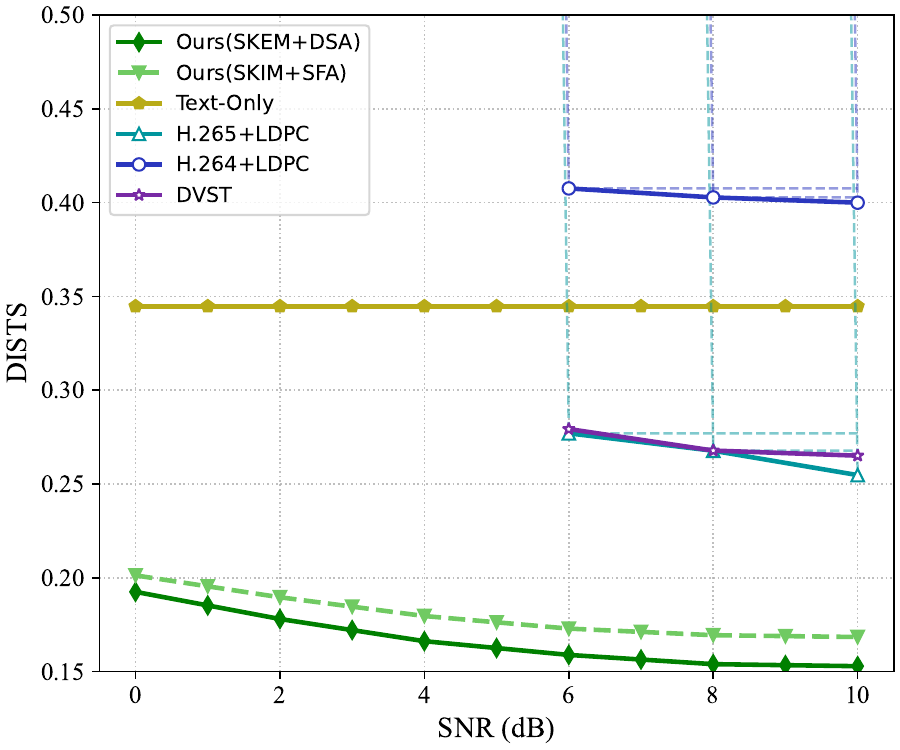}
        \caption{DISTS $\downarrow$}
        \label{fig:dists_vs_snr}
    \end{subfigure}
    \hspace{0.02\textwidth} %
    \begin{subfigure}[t]{0.22\textwidth}
        \centering
        \includegraphics[width=\textwidth]{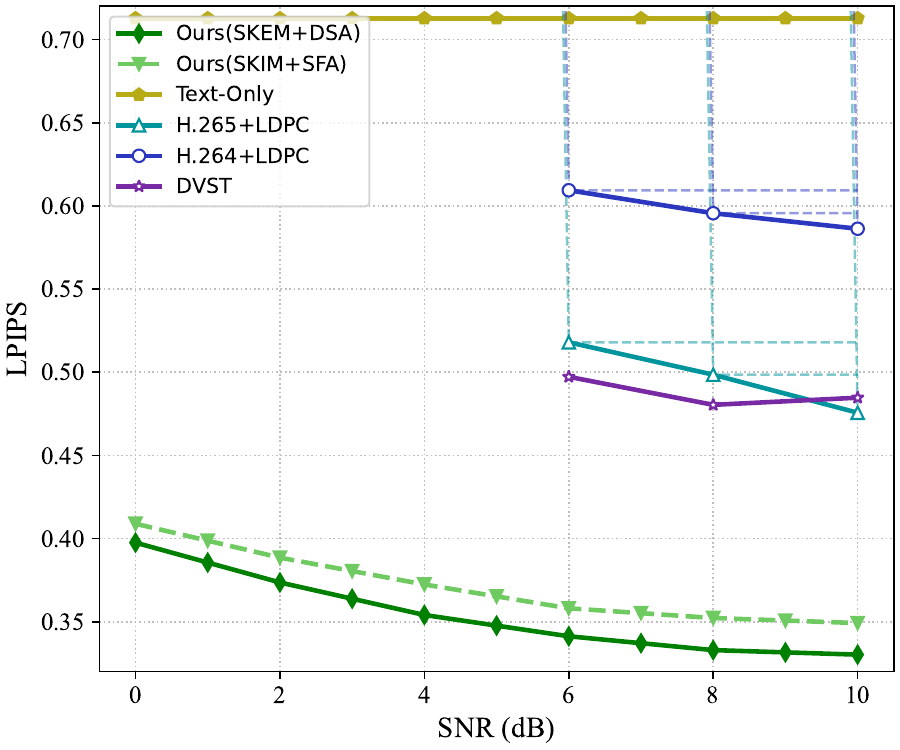}
        \caption{LPIPS $\downarrow$}
        \label{fig:lpips_vs_snr}
    \end{subfigure}

    \begin{subfigure}[t]{0.22\textwidth}
        \centering
        \includegraphics[width=\textwidth]{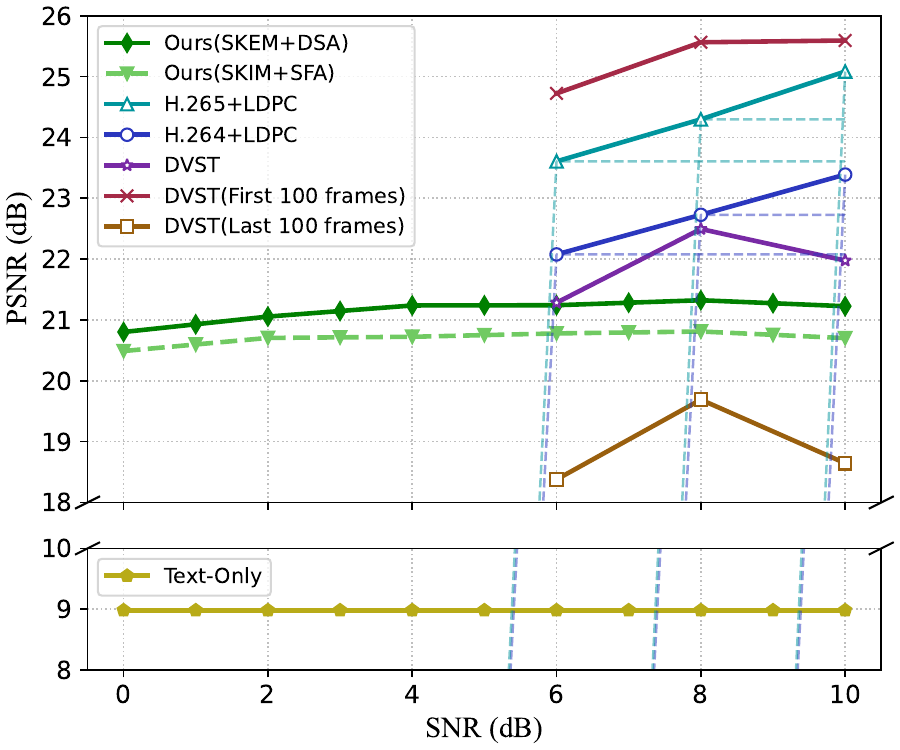}
        \caption{PSNR $\uparrow$}
        \label{fig:psnr_vs_snr}
    \end{subfigure}
    \hspace{0.02\textwidth} %
    \begin{subfigure}[t]{0.22\textwidth}
        \centering
        \includegraphics[width=\textwidth]{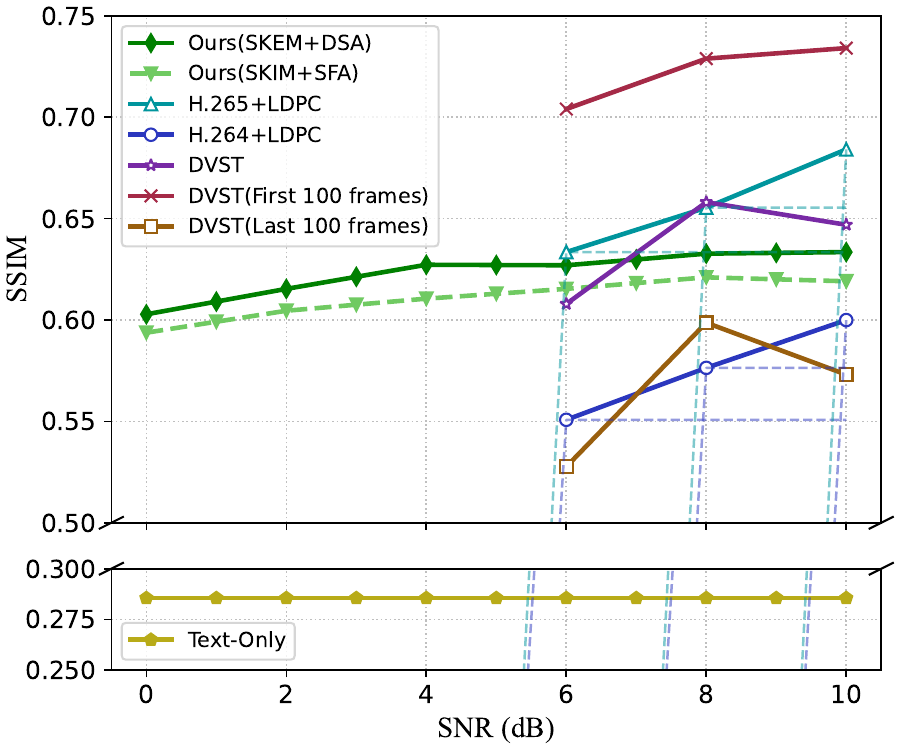}
        \caption{SSIM $\uparrow$}
        \label{fig:ssim_vs_snr}
    \end{subfigure}

    \caption{Performance of various quality metrics under different SNR levels.}
    \label{fig:metrics_vs_snr}
    \vspace{-5mm}
\end{figure}

In Fig.~\ref{fig:metrics_vs_snr}, the results across different metrics indicate the following key conclusions:
(1) The \textbf{Text-Only} scheme, despite operating at an extremely low CBR, demonstrates significant semantic fidelity, highlighting the effectiveness of the world model as a semantic decoder. It outperforms or closely matches other schemes in both CLIP and DISTS scores. 
(2) The \textbf{SKEM+DSA} scheme outperforms the \textbf{SKIM+SFA} baseline across all evaluation metrics, validating the effectiveness of our proposed PSSS metric. This enables more accurate keyframe selection and enhances semantic encoding. Since SKIM produces fixed-length segments (making DSA functionally equivalent to SFA in that setting), the full performance gain of SKEM+DSA over SKIM+SFA is attributable to SKEM's semantic-aware keyframe selection, while DSA serves as the necessary architectural enabler for handling variable-length segments.
(3) At ultra-low bitrates, when SNR $<$ 6 dB, traditional schemes including H.264/H.265+LDPC and DVST suffer from cliff effects and fail to transmit reliably. DVST, although a SemCom framework, exhibits severe color distortion indicative of semantic collapse. In contrast, our framework remains robust, maintaining semantic integrity under low-SNR conditions.

Specifically, Fig.~\ref{fig:clip_vs_snr}, \ref{fig:dists_vs_snr}, and \ref{fig:lpips_vs_snr} show results for deep-learning-based perceptual metrics. Our scheme achieves the satisfactory performance across all metrics, even with a lower CBR compared to other baselines. In contrast, Fig.~\ref{fig:psnr_vs_snr} and \ref{fig:ssim_vs_snr} depict results for traditional metrics, where our method does not achieve the highest scores. This is expected, as conventional codecs are optimized for PSNR and SSIM. However, qualitative visualizations (shown in Fig.~\ref{fig:visual_comparison}) reveal that, despite higher scores on traditional metrics, the reconstructed content is blurrier and semantically degraded. Our method preserves fine-grained details such as embedded watermarks, highlighting the limitations of traditional metrics in evaluating semantic quality.

\begin{table}[tbp]
\caption{Results of Video Description Task
\label{tab:video_description_results}}
\begin{center}
\scriptsize
\renewcommand{\arraystretch}{1.3}
\begin{tabular}{|>{\centering\arraybackslash}m{1.3cm}
                |>{\centering\arraybackslash}m{1.3cm}
                |>{\centering\arraybackslash}m{0.6cm}
                |>{\centering\arraybackslash}m{1.1cm}
                |>{\centering\arraybackslash}m{1.2cm}
                |>{\centering\arraybackslash}m{0.7cm}|}
\hline
\textbf{Metric} & \textbf{H.265+LDPC} & \textbf{DVST} & \textbf{SKIM+SFA} & \textbf{SKEM+DSA} & \textbf{Origin} \\
\hline
BLEU $\uparrow$       & 0.3444 & 0.3847 & 0.4317 & \textbf{0.4421} & 0.4556 \\
\hline
ROUGE-L $\uparrow$    & 0.2987 & 0.3144 & 0.3857 & \textbf{0.3992} & 0.4188 \\
\hline
BERT$\uparrow$ & 0.8830 & 0.8889 & 0.9023 & \textbf{0.9052} & 0.9076 \\
\hline
\end{tabular}
\end{center}
\vspace{-4mm}
\end{table}

Table~\ref{tab:video_description_results} presents results for the video description task. All metrics assess the similarity between the generated textual descriptions and human-annotated references. The \emph{Origin} column shows the performance with the original video. Our proposed SKEM+DSA outperforms all baselines and achieves nearly identical scores to those obtained from the original videos, demonstrating high semantic fidelity. The newly added SKIM+SFA baseline confirms that PSSS-guided keyframe selection (SKEM) provides additional gains over fixed-interval extraction, with BLEU improving from 0.4317 to 0.4421.
Table~\ref{tab:human_action_classification} presents results for the human action classification task, evaluated against the ground truth labels. Both SKIM+SFA and SKEM+DSA match the performance of the original video input, surpassing all other baselines. This indicates that high-level action recognition is robust to the keyframe selection strategy, while finer-grained tasks benefit more from semantic-aware selection (see Table~\ref{tab:depthany_results}).
Table~\ref{tab:depthany_results} shows the performance on the pixel-level monocular depth estimation task. All metrics assess the consistency of predictions on transmitted videos compared to the original videos. SKEM+DSA consistently outperforms SKIM+SFA across all metrics, with RMSE improving by 10.0\% (39.96 $\to$ 35.95), confirming that semantically selected keyframes better preserve fine-grained spatial details for machine perception tasks. Our approach scores lower on the AbsRel metric compared to DVST, due to the generative model used in the decoder, which is not explicitly constrained to preserve pixel-level fidelity.

\begin{table}[tbp]
\caption{Results of Human Action Classification Task
\label{tab:human_action_classification}}
\begin{center}
\scriptsize
\renewcommand{\arraystretch}{1.3}
\begin{tabular}{|>{\centering\arraybackslash}m{1.3cm}
                |>{\centering\arraybackslash}m{1.3cm}
                |>{\centering\arraybackslash}m{0.9cm}
                |>{\centering\arraybackslash}m{1.1cm}
                |>{\centering\arraybackslash}m{1.2cm}
                |>{\centering\arraybackslash}m{0.6cm}|}
\hline
\textbf{Metric} & \textbf{H.264+LDPC} & \textbf{DVST} & \textbf{SKIM+SFA} & \textbf{SKEM+DSA} & \textbf{Origin} \\
\hline
Top-1 Acc. & 64.29\% & 78.57\% & \textbf{85.71\%} & \textbf{85.71\%} & 85.71\% \\
\hline
Top-5 Acc. & 78.57\% & \textbf{92.86\%} & \textbf{92.86\%} & \textbf{92.86\%} & 92.86\% \\
\hline
\end{tabular}
\end{center}
\end{table}

\begin{table}[tp]
\caption{Results of Depth Estimation Task
\label{tab:depthany_results}}
\begin{center}
\scriptsize
\renewcommand{\arraystretch}{1.3}
\begin{tabular}{|>{\centering\arraybackslash}m{1.3cm}
                |>{\centering\arraybackslash}m{0.8cm}
                |>{\centering\arraybackslash}m{0.8cm}
                |>{\centering\arraybackslash}m{0.8cm}
                |>{\centering\arraybackslash}m{1.2cm}
                |>{\centering\arraybackslash}m{1.3cm}|}
\hline
\textbf{Method} & \text{D1 $\uparrow$} & \text{D2 $\uparrow$} & \text{D3 $\uparrow$} & \text{AbsRel $\downarrow$} & \text{RMSE $\downarrow$} \\
\hline
\textbf{H.265+LDPC} & 0.4445 & 0.6602 & 0.7775 & 1.1739 & 65.1621 \\
\hline
\textbf{DVST}       & 0.4083 & 0.6366 & 0.7600 & \textbf{0.7752} & 71.6045 \\
\hline
\textbf{SKIM+SFA} & 0.6803 & 0.8373 & 0.8964 & 1.4064 & 39.96 \\
\hline
\textbf{SKEM+DSA} & \textbf{0.7140} & \textbf{0.8646} & \textbf{0.9139} & 1.3285 & \textbf{35.95} \\
\hline
\end{tabular}
\end{center}
\vspace{-5mm}
\end{table}

\subsection{Computational complexity analysis}
We report the end-to-end latency of the LGVSC pipeline measured on our testbed. At the transmitter, SKEM preprocessing (InternVL captioning + PSSS scoring) takes approximately $2{,}444$~s per minute of input video on an NVIDIA RTX~4090 (165.2~BF16 TFLOPS), with each PSSS frame-pair evaluation requiring $8.77$~s (of which 95.7\% is consumed by Round-1 caption generation). NTSCC keyframe encoding adds only 46.5~ms per frame. At the receiver, Open-Sora video reconstruction runs at approximately 24.5~s per generated second of video on an RTX~4080 SUPER.

Projecting the measured BF16 latencies to data-center GPUs using peak TFLOPS scaling, an NVIDIA B200 SXM (2250~BF16 TFLOPS) would reduce PSSS scoring to around $0.64$~s/pair and SKEM preprocessing to around $179$~s/min of video, a $13.6\times$ speedup over the RTX~4090. On H100/H200 SXM GPUs (989.4~BF16 TFLOPS) already in widespread deployment, the projected speedup is $6.0\times$. Furthermore, the rapid evolution of efficient diffusion inference is independently reducing receiver-side reconstruction cost. LGVSC thus benefits from two converging trends, namely hardware scaling and software architectural advances, which together {may progressively narrow the gap to near-real-time operation. Nevertheless, the present pipeline is best positioned for offline or edge-cloud-assisted scenarios rather than real-time wireless transmission}.

\section{Conclusions}

In this work, we propose LGVSC, a large-model-driven generative video semantic communication framework designed for high-fidelity video transmission under extremely low-bitrate conditions. A novel probability-based semantic similarity score is introduced to enable fine-grained semantic evaluation and guide adaptive keyframe extraction through the SKEM module. The proposed SKEM+DSA scheme achieves state-of-the-art performance in both semantic and perceptual quality metrics {while operating at a channel bandwidth ratio on the order of $10^{-4}$ to $10^{-3}$}. LGVSC effectively balances human perceptual needs and downstream machine-task requirements, demonstrating its significant potential for practical deployment in future bandwidth-constrained IoE scenarios. The current LGVSC framework targets general-purpose video semantic transmission {in offline or edge-cloud-assisted scenarios. Real-time transmission and} safety-critical applications such as autonomous driving and medical video transmission {both} require further dedicated investigation.


\FloatBarrier
\bibliographystyle{ieeetr}
\bibliography{FL_Ref}

\begin{thebibliography}{10}

\bibitem{ying2026specialist}
K.~Ying, Z.~Gao, T.~Yang, J.~Zhang, X.~Cheng, T.~Q. Quek, and H.~V. Poor, ``From specialist to large models: A paradigm evolution towards semantic-aware mimo,'' {\em IEEE Communications Magazine}, pp.~1--8, 2026.

\bibitem{yinYinhangGenerativeVideoSemantic2025}
H.~Yin, L.~Qiao, Y.~Ma, S.~Sun, K.~Li, Z.~Gao, and D.~Niyato, ``Generative video semantic communication via multimodal semantic fusion with large model,'' {\em {IEEE} Trans. Veh. Technol.}, vol.~75, no.~1, pp.~1701--1706, 2026.

\bibitem{wangDVSTWirelessDeepVideo2023}
S.~Wang, J.~Dai, Z.~Liang, K.~Niu, Z.~Si, C.~Dong, X.~Qin, and P.~Zhang, ``Wireless deep video semantic transmission,'' {\em {IEEE} J. Select. Areas Commun.}, vol.~41, pp.~214--229, Jan. 2023.

\bibitem{tongMultimodalSemanticCommunication2025}
H.~Tong, H.~Li, H.~Du, Z.~Yang, C.~Yin, and D.~Niyato, ``Multimodal semantic communication for generative audio-driven video conferencing,'' {\em IEEE Wireless Commun. Lett.}, vol.~14, pp.~93--97, Jan. 2025.

\bibitem{guoVideoQASCAdaptiveSemantic2025a}
J.~Guo, W.~Chen, Y.~Sun, J.~Xu, and B.~Ai, ``Videoqa-sc: Adaptive semantic communication for video question answering,'' {\em {IEEE} J. Select. Areas Commun.}, vol.~43, pp.~2462--2477, July 2025.

\bibitem{qiaoTokenCommunicationsLarge2025}
L.~{Qiao}, M.~B. {Mashhadi}, Z.~{Gao}, R.~{Tafazolli}, M.~{Bennis}, and D.~{Niyato}, ``Token communications: A large model-driven framework for cross-modal context-aware semantic communications,'' {\em {IEEE} Wireless Commun.}, vol.~32, pp.~80--88, Jan. 2025.

\bibitem{li2025large}
Z.~Li, Z.~Gao, X.~Liu, Z.~Wang, X.~Zhou, L.~Liu, Y.~Wu, W.~Feng, and Y.~Huang, ``Large model enabled embodied intelligence for 6g integrated perception, communication, and computation network,'' {\em arXiv preprint arXiv:2512.15109}, 2025.

\bibitem{liu2024sorareviewbackgroundtechnology}
Y.~Liu, K.~Zhang, Y.~Li, Z.~Yan, C.~Gao, R.~Chen, Z.~Yuan, Y.~Huang, H.~Sun, J.~Gao, L.~He, and L.~Sun, ``Sora: A review on background, technology, limitations, and opportunities of large vision models,'' {\em arXiv preprint arXiv:2402.17177}, 2024.

\bibitem{qiaoLatencyAwareGenerativeSemantic2024}
L.~Qiao, M.~B. Mashhadi, Z.~Gao, C.~H. Foh, P.~Xiao, and M.~Bennis, ``Latency-aware generative semantic communications with pre-trained diffusion models,'' {\em IEEE Wireless Commun. Lett.}, vol.~13, pp.~2652--2656, Oct. 2024.

\bibitem{NTSCC}
J.~Dai, S.~Wang, K.~Tan, Z.~Si, X.~Qin, K.~Niu, and P.~Zhang, ``Nonlinear transform source-channel coding for semantic communications,'' {\em {IEEE} J. Select. Areas Commun.}, vol.~40, no.~8, pp.~2300--2316, 2022.

\bibitem{TIFA}
Y.~Hu {\em et~al.}, ``Tifa: Accurate and interpretable text-to-image faithfulness evaluation with question answering,'' in {\em Proc. IEEE Int. Conf. Comput. Vis. (ICCV)}, pp.~20349--20360, 2023.

\bibitem{opticalflow}
H.~Xu, J.~Zhang, J.~Cai, H.~Rezatofighi, F.~Yu, D.~Tao, and A.~Geiger, ``Unifying flow, stereo and depth estimation,'' {\em IEEE Trans. Pattern Anal. Mach. Intell.}, vol.~45, no.~11, pp.~13941--13958, 2023.

\bibitem{chen2024internvlscalingvisionfoundation}
Z.~Chen {\em et~al.}, ``Internvl: Scaling up vision foundation models and aligning for generic visual-linguistic tasks,'' in {\em Proc. IEEE Conf. Comput. Vis. Pattern Recognit. (CVPR)}, pp.~24185--24198, June 2024.

\bibitem{opensora}
Z.~Zheng, X.~Peng, T.~Yang, C.~Shen, S.~Li, H.~Liu, Y.~Zhou, T.~Li, and Y.~You, ``Open-sora: Democratizing efficient video production for all,'' {\em arXiv preprint arXiv:2412.20404}, 2024.

\bibitem{WebVid}
M.~Bain, A.~Nagrani, G.~Varol, and A.~Zisserman, ``Frozen in time: A joint video and image encoder for end-to-end retrieval,'' in {\em Proc. IEEE Int. Conf. Comput. Vis. (ICCV)}, pp.~1728--1738, October 2021.

\bibitem{Kinetics400}
J.~Carreira and A.~Zisserman, ``Quo vadis, action recognition? a new model and the kinetics dataset,'' in {\em Proc. IEEE Conf. Comput. Vis. Pattern Recognit. (CVPR)}, pp.~6299--6308, July 2017.

\bibitem{h264}
{ITU-T and ISO/IEC JTC 1}, ``Advanced video coding for generic audiovisual services.'' ITU-T Recommendation H.264 and ISO/IEC 14496-10, 2010.

\bibitem{h265}
G.~J. Sullivan, J.-R. Ohm, W.-J. Han, and T.~Wiegand, ``Overview of the high efficiency video coding (hevc) standard,'' {\em IEEE Trans. Circuit Syst. Video Technol.}, vol.~22, no.~12, pp.~1649--1668, 2012.

\bibitem{xu2024pllavaparameterfreellava}
L.~Xu, Y.~Zhao, D.~Zhou, Z.~Lin, S.~K. Ng, and J.~Feng, ``Pllava: Parameter-free llava extension from images to videos for video dense captioning,'' 2024.

\bibitem{TimeSformer}
G.~Bertasius, H.~Wang, and L.~Torresani, ``Is space-time attention all you need for video understanding?,'' in {\em Proc. Int. Conf. Mach. Learn. (ICML)}, July 2021.

\bibitem{depth_anything_v2}
L.~Yang, B.~Kang, Z.~Huang, Z.~Zhao, X.~Xu, J.~Feng, and H.~Zhao, ``Depth anything v2,'' {\em Proc. Adv. Neural Inf. Process. Syst. (NeurIPS)}, vol.~37, pp.~21875--21911, 2024.

\bibitem{zhangbertscore}
T.~Zhang, V.~Kishore, F.~Wu, K.~Q. Weinberger, and Y.~Artzi, ``Bertscore: Evaluating text generation with bert,'' in {\em Proc. Int. Conf. Learn. Representations (ICLR)}, 2020.

\end{thebibliography}

\clearpage
\appendices

\section{PSSS Validation: Case Study Comparing PSSS and CLIP}
\label{appendix:case_study}

A core advantage of PSSS over CLIP-based similarity is its use of language-level semantic descriptions rather than global visual features, enabling it to detect semantic transitions that CLIP cannot. To illustrate this, we curate two categories of challenging frame pairs from the 55-video WebVid test set where PSSS and CLIP \emph{disagree}, and show that PSSS aligns with human judgment in each case. The CLIP threshold for ``sufficiently different'' is set at cosine similarity~$<0.95$, following standard contrastive practice.

\begin{figure*}[bp]
\centering
\begin{subfigure}[t]{0.48\textwidth}
    \includegraphics[width=\textwidth]{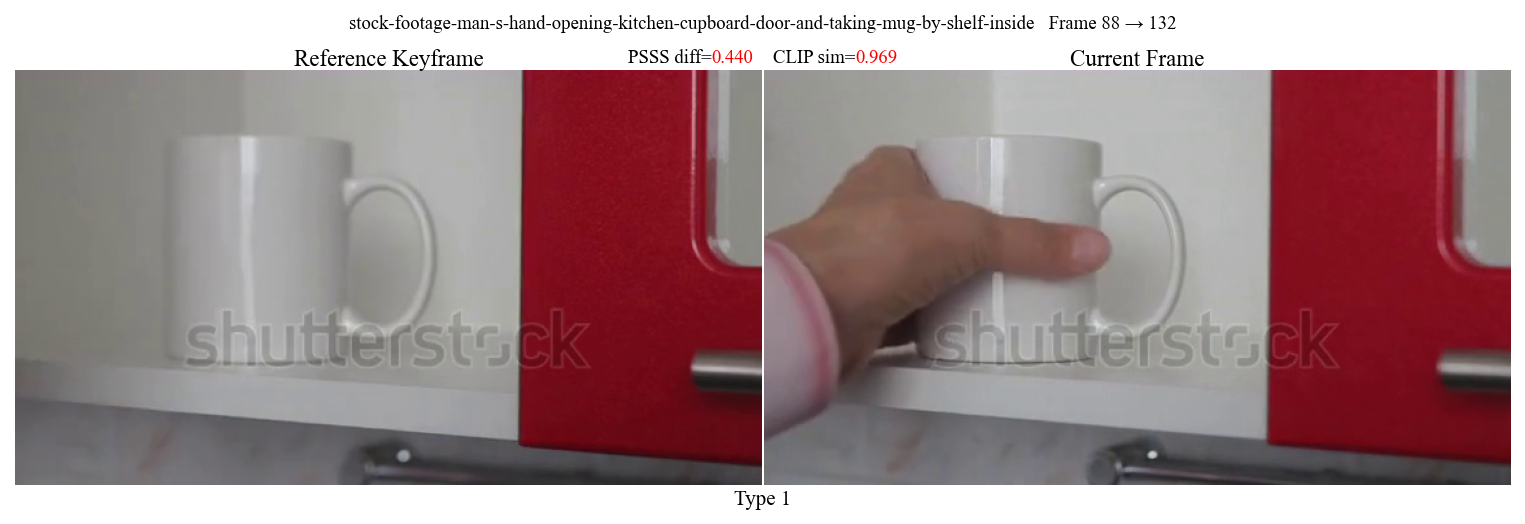}
    \caption{Kitchen cupboard~(1).}
\end{subfigure}
\hfill
\begin{subfigure}[t]{0.48\textwidth}
    \includegraphics[width=\textwidth]{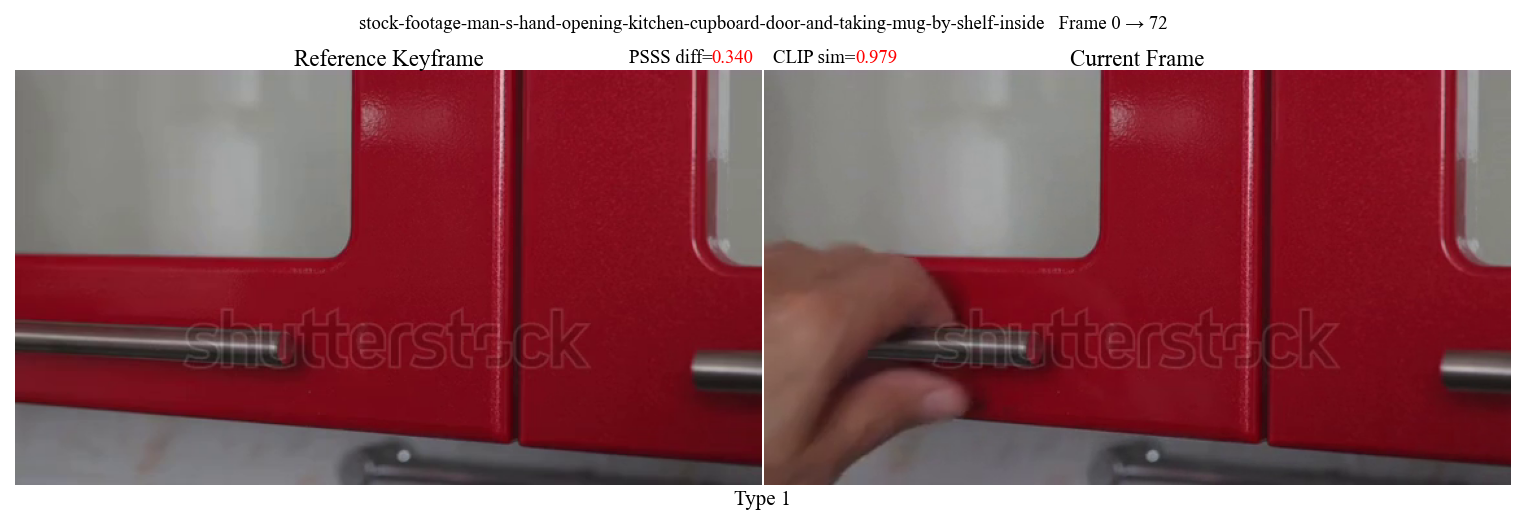}
    \caption{Kitchen cupboard~(2).}
\end{subfigure}
\\[4pt]
\begin{subfigure}[t]{0.48\textwidth}
    \includegraphics[width=\textwidth]{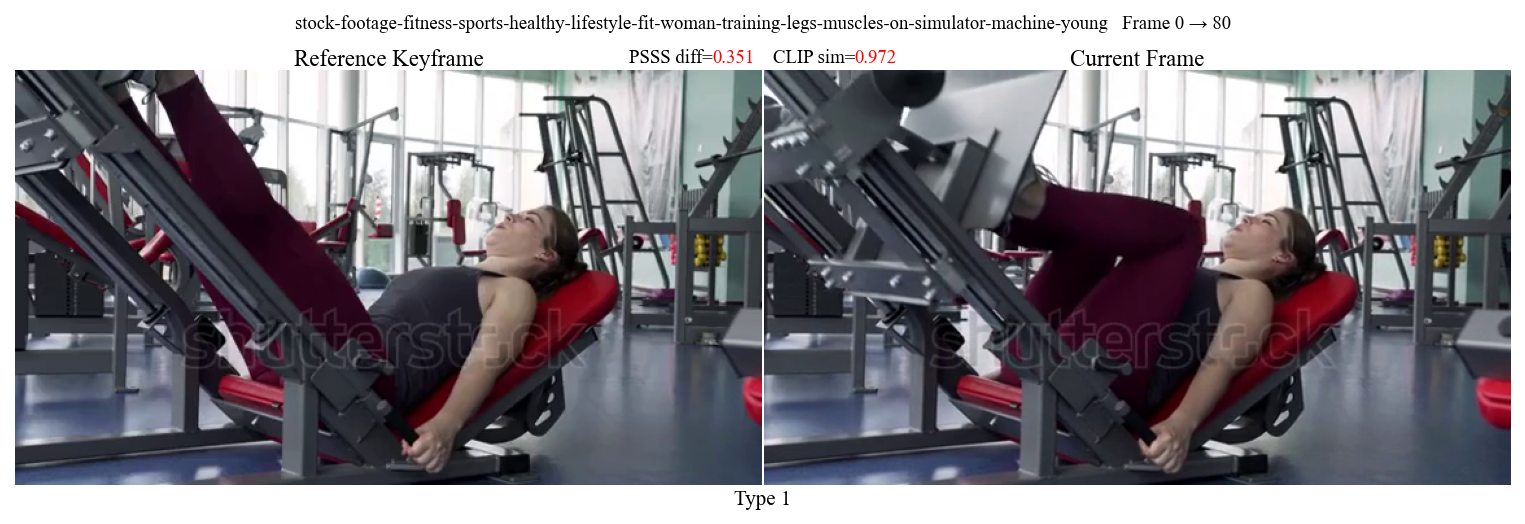}
    \caption{Leg-press machine.}
\end{subfigure}
\hfill
\begin{subfigure}[t]{0.48\textwidth}
    \includegraphics[width=\textwidth]{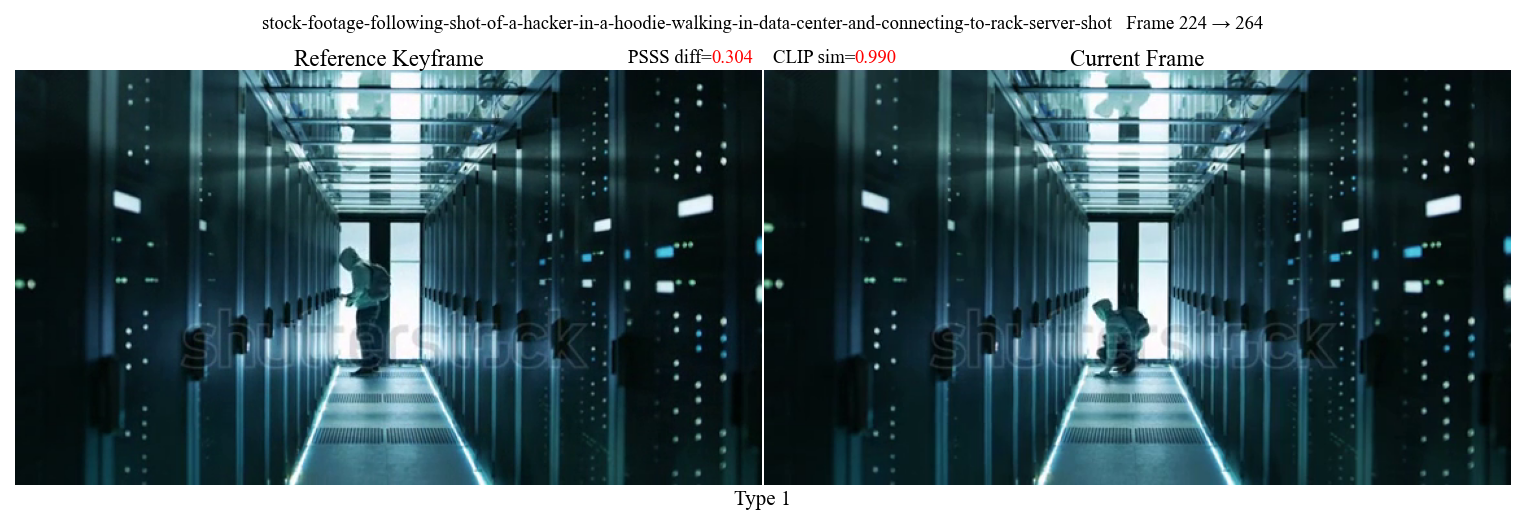}
    \caption{Data center.}
\end{subfigure}
\caption{Type~1 case study: frame pairs with high CLIP similarity ($\geq0.97$) yet genuine semantic transitions. PSSS correctly triggers keyframe insertion ($S_\mathrm{rel}>\eta_\mathrm{th}$); a CLIP-based threshold would not.}
\label{fig:app_type1}
\end{figure*}

\textbf{Type~1: Visually similar but semantically changing.}
These frame pairs have CLIP cosine similarity~$\geq0.96$, which would suppress keyframe insertion under a CLIP threshold. Yet each pair contains a genuine semantic transition (an action onset, a new actor, or a significant positional change) that PSSS correctly detects ($S_\mathrm{rel}>\eta_\mathrm{th}=0.35$). Because CLIP embeddings capture global appearance statistics, they are insensitive to localized semantic transitions when the background is nearly identical. PSSS, operating on language-level descriptions, correctly identifies these changes regardless of background appearance. Representative examples are shown in Fig.~\ref{fig:app_type1}.

\begin{figure*}[bp]
\centering
\begin{subfigure}[t]{0.48\textwidth}
    \includegraphics[width=\textwidth]{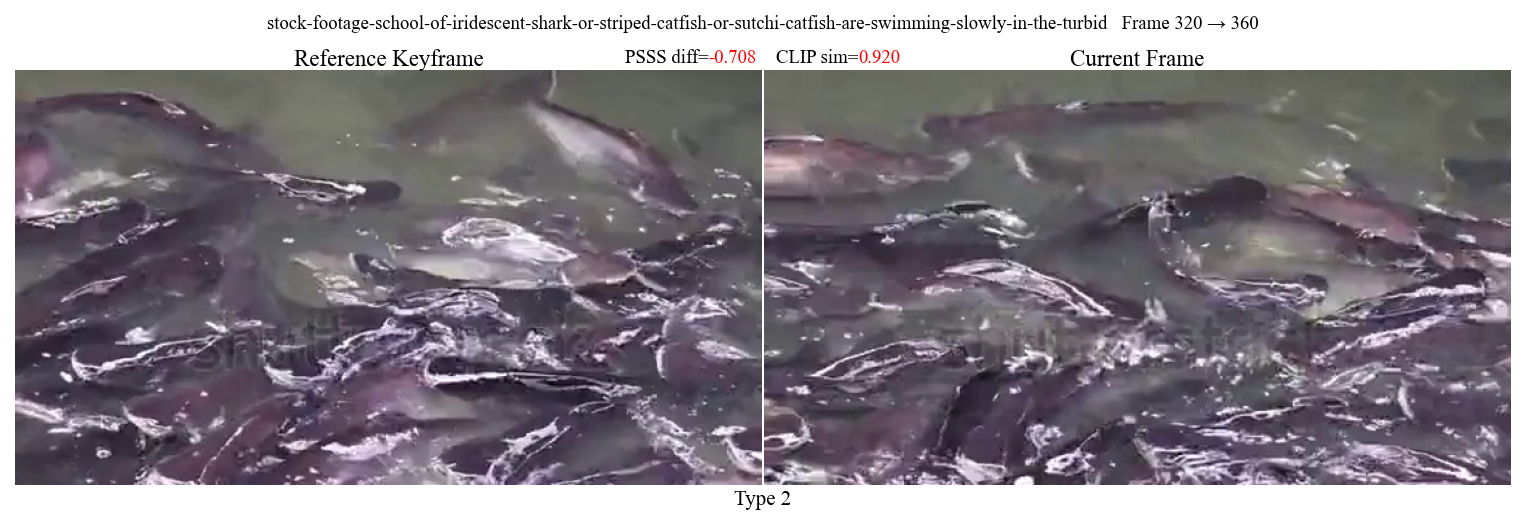}
    \caption{Catfish.}
\end{subfigure}
\hfill
\begin{subfigure}[t]{0.48\textwidth}
    \includegraphics[width=\textwidth]{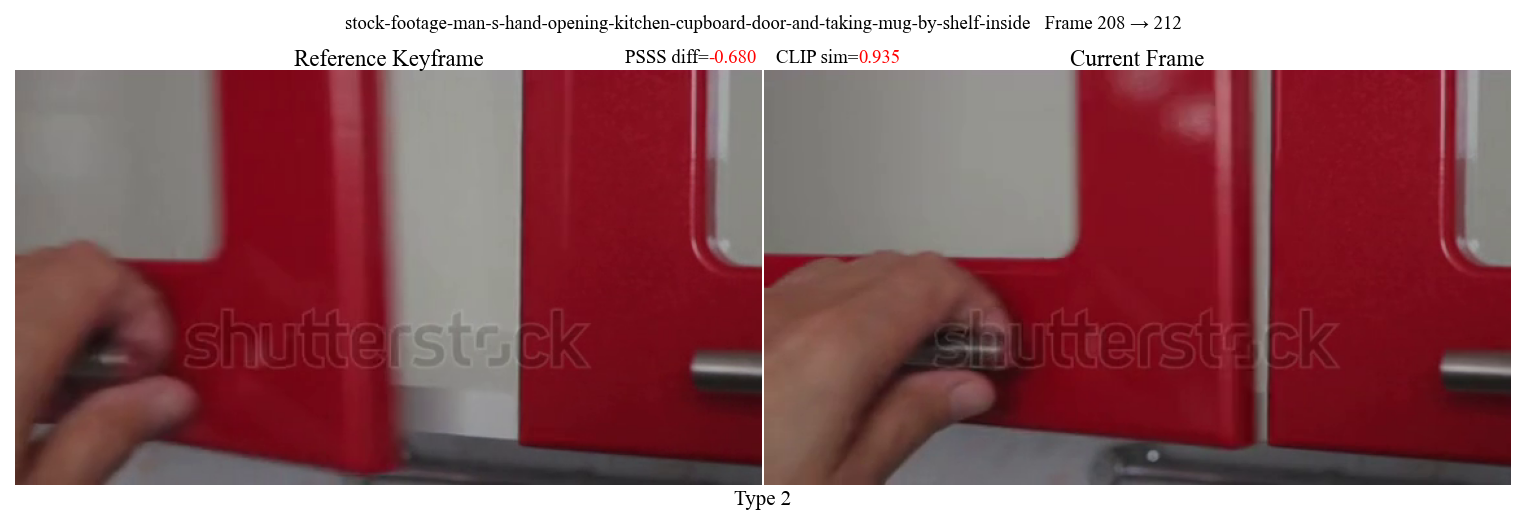}
    \caption{Kitchen cupboard.}
\end{subfigure}
\\[4pt]
\begin{subfigure}[t]{0.48\textwidth}
    \includegraphics[width=\textwidth]{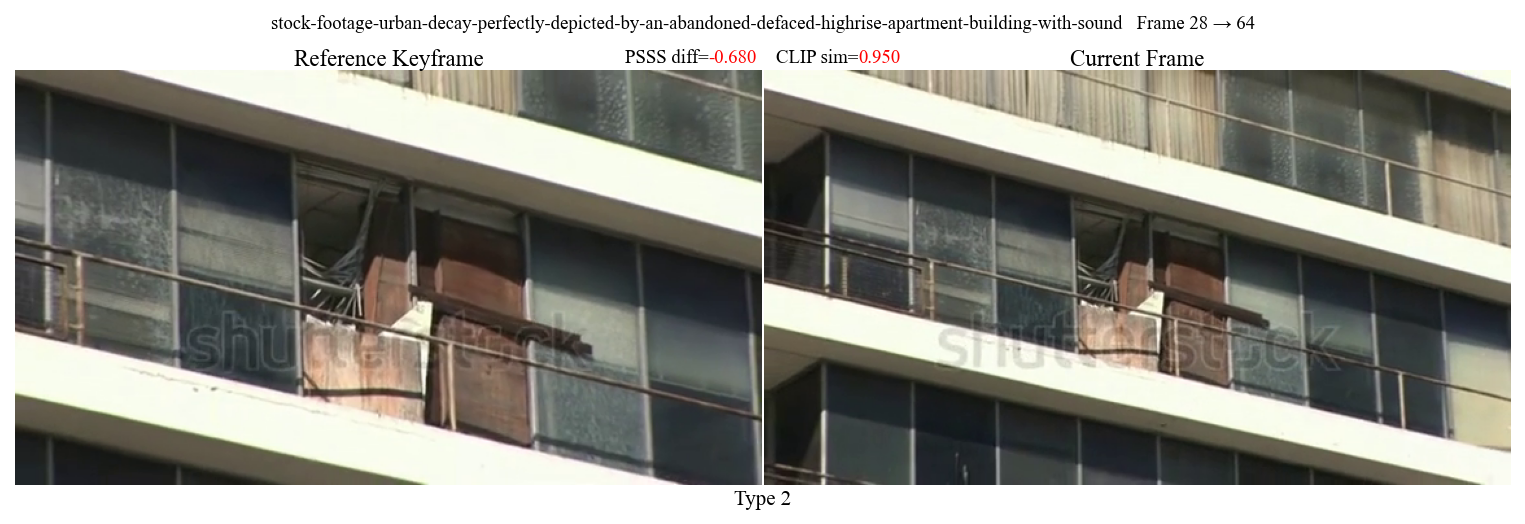}
    \caption{Urban building.}
\end{subfigure}
\hfill
\begin{subfigure}[t]{0.48\textwidth}
    \includegraphics[width=\textwidth]{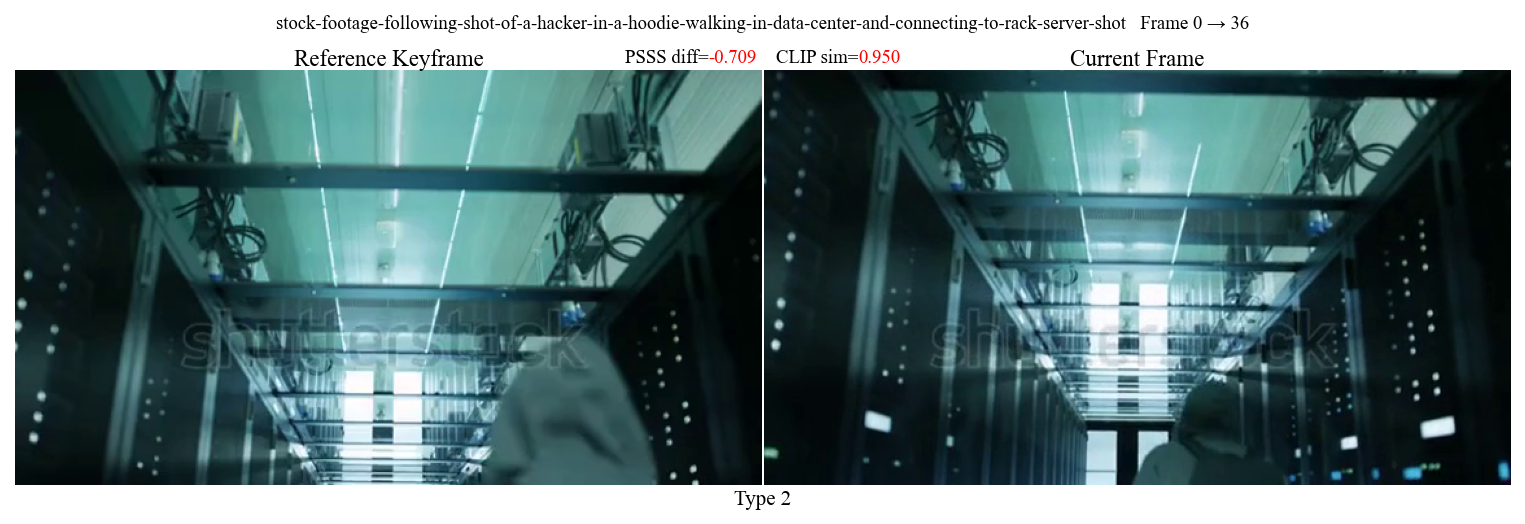}
    \caption{Data center.}
\end{subfigure}
\caption{Type~2 case study: frame pairs with moderate CLIP similarity ($0.92$--$0.95$) yet stable semantic content. PSSS signals continuity ($S_\mathrm{rel}\ll 0$) and suppresses redundant keyframe insertions that a CLIP threshold would trigger.}
\label{fig:app_type2}
\end{figure*}

\textbf{Type~2: Semantically continuous with visual variation.}
These frame pairs have moderate CLIP similarity ($0.92$--$0.95$), a range where a CLIP threshold might trigger spurious keyframe insertion. Yet the semantic content is stable (continuous action, same scene, same actors) and PSSS decisively signals continuity ($S_\mathrm{rel}\ll 0$). The visual variation arises from low-level factors such as motion blur, camera pan, or subject motion, none of which constitute a semantic state change. Representative examples are shown in Fig.~\ref{fig:app_type2}.

A particularly illustrative cross-case comparison arises between Kitchen cupboard~(1) in Type~1 (Fig.~\ref{fig:app_type1}(a)) and Kitchen cupboard in Type~2 (Fig.~\ref{fig:app_type2}(b)): both depict the same cupboard-opening action, yet PSSS correctly selects the moment the action \emph{begins} as a keyframe while decisively suppressing the intermediate continuous process as redundant. This directly reflects the semantic objective of keyframe selection: capturing \emph{state transitions} rather than uniformly sampling visual frames.

\section{Semantic Focus Flexibility of PSSS}
\label{appendix:semantic_focus}

Human perception of semantic similarity is context-dependent: two frames may be similar \emph{from the perspective of the background} yet different \emph{from the perspective of the foreground action}. PSSS captures this via the $\langle\text{Semantic Focus}\rangle$ parameter in its semantic similarity evaluation prompt. This appendix details the three focus variants and illustrates their behavior on five representative frame pairs.

\textbf{Semantic similarity evaluation prompts.}
The three focus variants differ only in the semantic aspect they ask the model to attend to:
\begin{itemize}
  \item \textbf{General:} \emph{``Compare the two descriptions of the images you have given. Focus on the semantic similarity of the images: the positions of key objects in the scene, any objects that have appeared or disappeared and the extent of changes in the background environment. Determine if these aspects depict the exact same scene. Only respond with `yes' if they match, otherwise respond with `no'.''}

  \item \textbf{Main Object:} \emph{``Compare the two descriptions of the images you have given. Focus only on the actions and movements: what people or objects are doing, their poses, gestures, and motion. Ignore any differences in the background or environment. Determine if the actions depicted are semantically the same. Only respond with `yes' if the actions match, otherwise respond with `no'.''}

  \item \textbf{Background:} \emph{``Compare the two descriptions of the images you have given. Focus only on the background environment: the setting, scene type, ambient lighting, and static surroundings. Ignore any people, objects, or their actions in the foreground. Determine if the background environments are semantically the same. Only respond with `yes' if the backgrounds match, otherwise respond with `no'.''}
\end{itemize}

\begin{figure*}[bp]
\centering
\begin{subfigure}[t]{0.48\textwidth}
    \includegraphics[width=\textwidth]{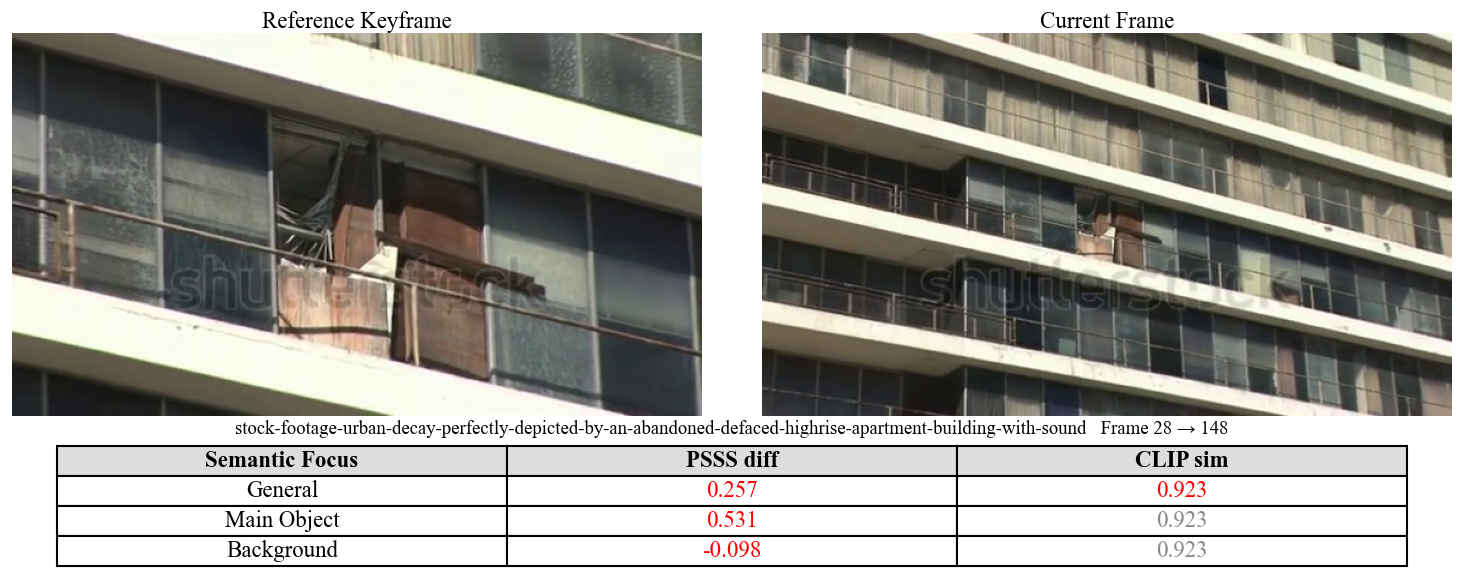}
    \caption{Urban building. $S_\mathrm{rel}^\mathrm{main}{=}{+}0.531$,\; $S_\mathrm{rel}^\mathrm{bg}{=}{-}0.098$.}
\end{subfigure}
\hfill
\begin{subfigure}[t]{0.48\textwidth}
    \includegraphics[width=\textwidth]{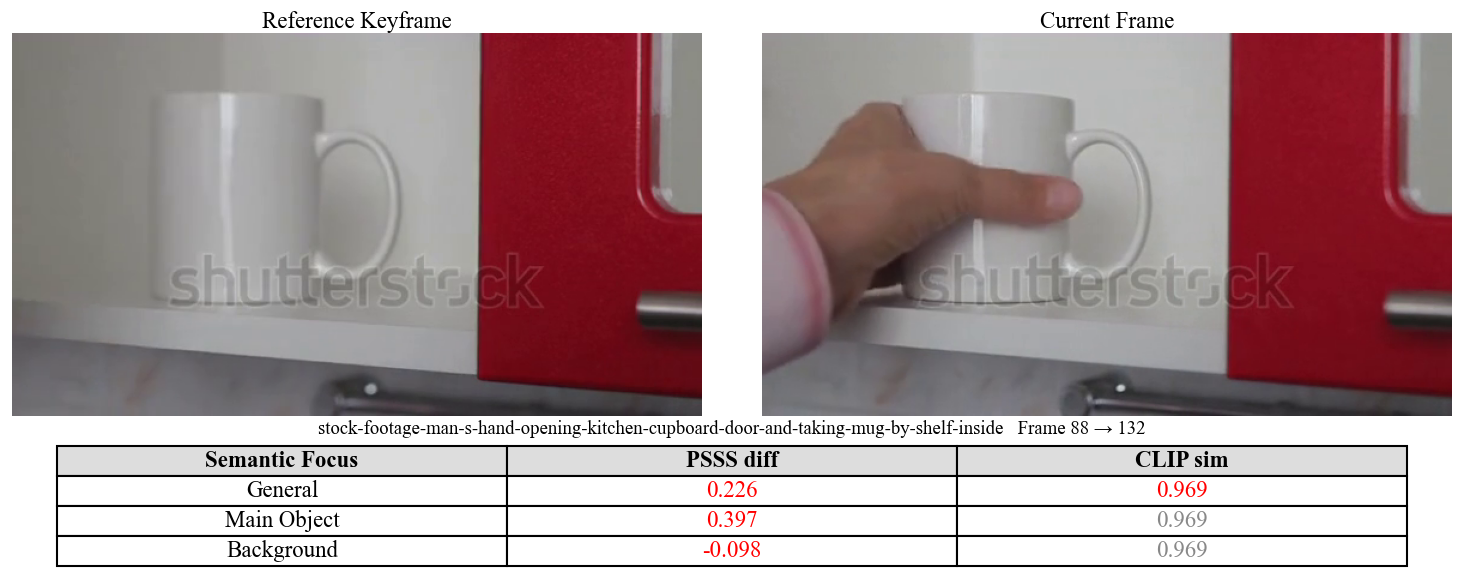}
    \caption{Kitchen cupboard. $S_\mathrm{rel}^\mathrm{main}{=}{+}0.397$,\; $S_\mathrm{rel}^\mathrm{bg}{=}{-}0.098$.}
\end{subfigure}
\\[4pt]
\begin{subfigure}[t]{0.48\textwidth}
    \includegraphics[width=\textwidth]{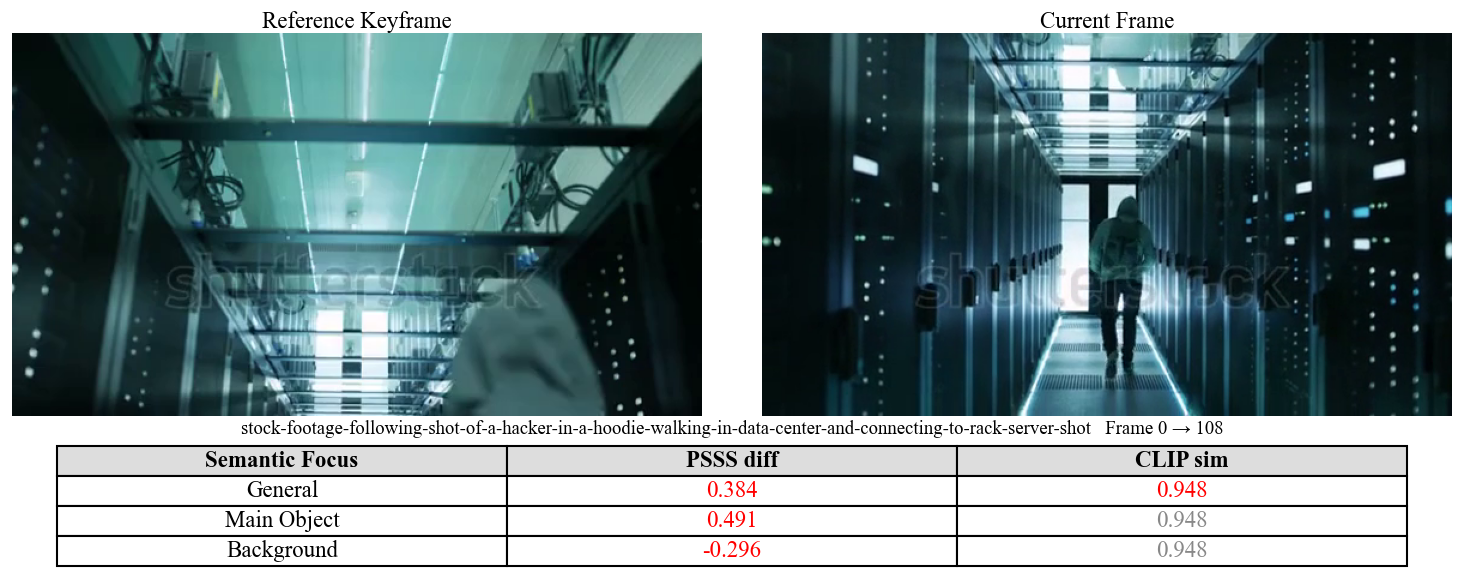}
    \caption{Data center. $S_\mathrm{rel}^\mathrm{main}{=}{+}0.491$,\; $S_\mathrm{rel}^\mathrm{bg}{=}{-}0.296$.}
\end{subfigure}
\hfill
\begin{subfigure}[t]{0.48\textwidth}
    \includegraphics[width=\textwidth]{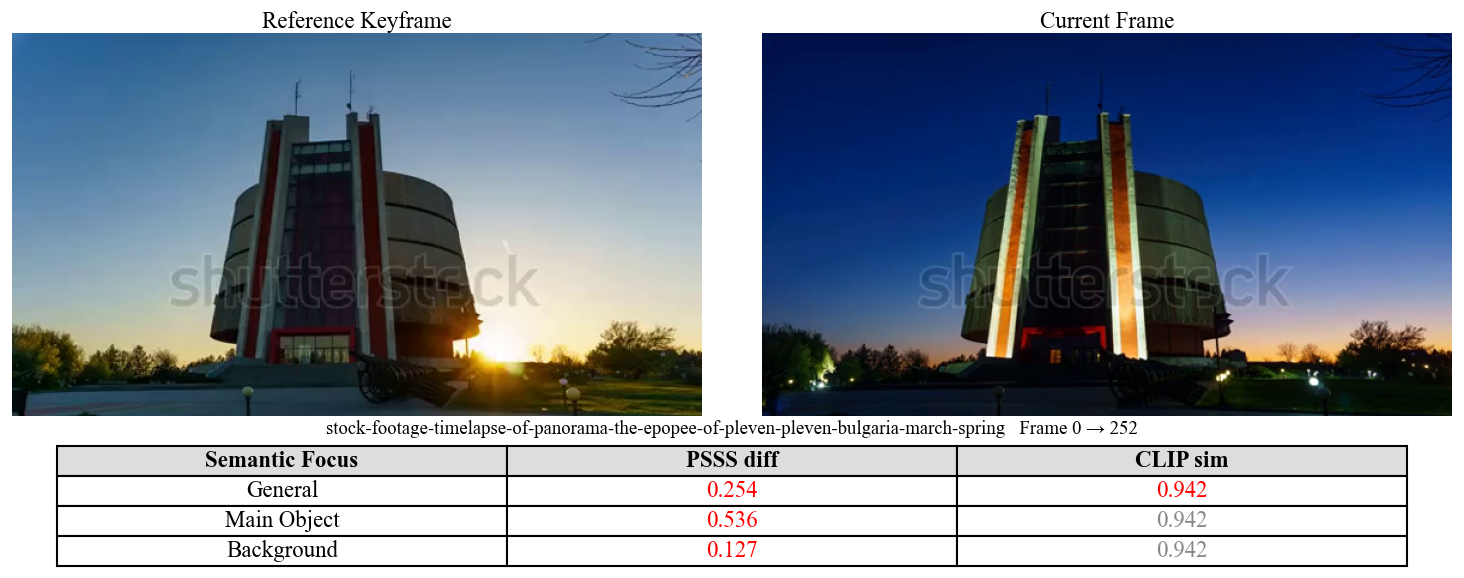}
    \caption{Pleven timelapse. $S_\mathrm{rel}^\mathrm{main}{=}{+}0.536$,\; $S_\mathrm{rel}^\mathrm{bg}{=}{+}0.127$.}
\end{subfigure}
\\[4pt]
\hfil
\begin{subfigure}[t]{0.48\textwidth}
    \includegraphics[width=\textwidth]{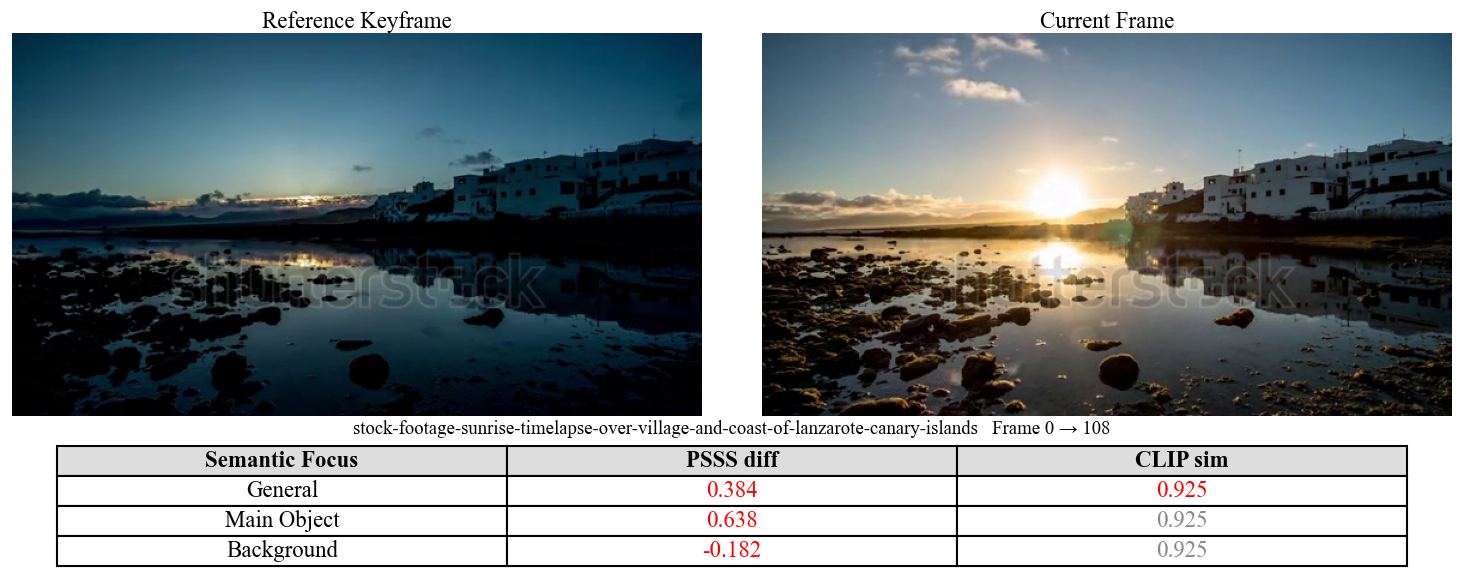}
    \caption{Lanzarote sunrise. $S_\mathrm{rel}^\mathrm{main}{=}{+}0.638$,\; $S_\mathrm{rel}^\mathrm{bg}{=}{-}0.182$.}
\end{subfigure}
\hfil
\caption{PSSS Semantic Focus flexibility. Switching the $\langle$Semantic Focus$\rangle$ from ``main object'' to ``background'' changes $S_\mathrm{rel}$ substantially across all five pairs (a larger value indicates greater semantic divergence), while CLIP returns a fixed scalar in the range $0.923$--$0.969$ regardless of focus.}
\label{fig:app_flex}
\end{figure*}

Despite sharing the same sentence structure, the three prompts produce substantially different $S_\mathrm{rel}$ scores because they direct the model's attention to different semantic aspects, as shown in Fig.~\ref{fig:app_flex}.

\textbf{Per-case analysis.}
In all five examples, the PSSS scores under each focus align with human perceptual judgment:
\begin{itemize}
  \item \textbf{Urban building (Fig.~\ref{fig:app_flex}(a)):} The camera pulls back from a close-up of a damaged balcony to a wide shot of the derelict high-rise facade. The visual \emph{subject} shifts substantially (high main-object score $+0.531$), while the underlying building structure in the background is unchanged (near-zero negative background score $-0.098$).

  \item \textbf{Kitchen cupboard (Fig.~\ref{fig:app_flex}(b)):} A hand reaches into the cupboard and grasps a mug. The main-object focus detects the action change (score $+0.397$), while the background focus correctly identifies that the cupboard interior and kitchen setting are unchanged ($-0.098$).

  \item \textbf{Data center (Fig.~\ref{fig:app_flex}(c)):} Despite a substantial camera angle change between the two frames, PSSS correctly recognizes that the \emph{background} (server-rack corridor) is semantically unchanged (strong negative score $-0.296$), while the \emph{person's action} has changed (score $+0.491$). A pixel-level metric such as CLIP would be confused by the viewpoint shift; PSSS, operating on language descriptions, is viewpoint-invariant.

  \item \textbf{Pleven timelapse (Fig.~\ref{fig:app_flex}(d)):} Building lights turn on as the sun sets, detected as a main-object change ($+0.536$). The gradual sky transition is a mild background change, captured as a slightly positive background score ($+0.127$) rather than a strong one, consistent with human perception.

  \item \textbf{Lanzarote sunrise (Fig.~\ref{fig:app_flex}(e)):} The sun rises visibly above the horizon, representing the clearest main-object change among the five examples (highest score, $+0.638$). The coastal village and sea remain structurally unchanged (negative background score $-0.182$).
\end{itemize}
CLIP, by contrast, returns cosine similarities in the range $0.923$--$0.969$ for all five pairs regardless of which semantic aspect is queried, and cannot distinguish foreground change from background change. This task-specific controllability is unique to PSSS's prompt-based design: practitioners can steer keyframe selection toward the semantically relevant dimension for their downstream task (e.g., foreground action for action recognition, background scene for environment monitoring) without any model retraining.

\section{Stability of the Relative Probability Formulation}
\label{appendix:robustness}

Section~\ref{subsec:psss} argues theoretically that $S_\mathrm{rel}=P(\text{``No''})-P(\text{``Yes''})$ is more stable than using the absolute token probability $P(\text{``No''})$ directly, because common-mode shifts in model confidence cancel in the difference. Here we provide a controlled experiment to verify this claim.

\textbf{Setup.}
We run SKEM on 21 WebVid validation videos, keeping the model (InternVL2-8B), prompt, and all other settings identical, varying only the scoring rule:
\begin{itemize}
  \item \textbf{Relative ($\delta$):} $S_\mathrm{rel} = P(\text{``No''})-P(\text{``Yes''}) > 0.35$
  \item \textbf{Absolute (abs):} $P(\text{``No''}) > 0.4$
\end{itemize}

Table~\ref{tab:delta_ablation} reports the resulting keyframe count statistics per video.

\begin{table}[htbp]
\caption{Keyframe count statistics: relative ($\delta$) vs.\ absolute probability threshold on 21 WebVid validation videos.
\label{tab:delta_ablation}}
\begin{center}
\scriptsize
\renewcommand{\arraystretch}{1.3}
\begin{tabular}{|>{\centering\arraybackslash}m{2.6cm}
                |>{\centering\arraybackslash}m{0.75cm}
                |>{\centering\arraybackslash}m{0.75cm}
                |>{\centering\arraybackslash}m{0.55cm}
                |>{\centering\arraybackslash}m{0.55cm}
                |>{\centering\arraybackslash}m{1.3cm}|}
\hline
\textbf{Scoring Rule} & \textbf{Mean} & \textbf{Std} & \textbf{Min} & \textbf{Max} & \textbf{CV} \\
\hline
$\delta$: $P(\text{No}){-}P(\text{Yes}){>}0.35$ & 4.10 & 2.70 & 2 & 12 & \textbf{0.659} \\
\hline
Abs: $P(\text{No}){>}0.4$ & 9.48 & 10.08 & 2 & 38 & \textbf{1.064} \\
\hline
\end{tabular}
\end{center}
\end{table}

\textbf{Results and interpretation.}
The $\delta$ formula reduces the coefficient of variation (CV\,$=$\,Std/Mean) by 38\% (from $1.064$ to $0.659$) and cuts the maximum keyframe count from 38 to 12. Furthermore, 12 out of 21 videos (57\%) differ by three or more keyframes between the two rules, confirming that the gap reflects a systematic structural difference rather than random noise.

The mechanism is as follows. The PSSS prompt constrains the model to a binary yes/no format, concentrating probability mass onto a small token set. When the model is uncertain about a frame pair, for instance when two frames have nearly identical visual content, probability mass spreads across both tokens simultaneously, causing both $P(\text{``No''})$ and $P(\text{``Yes''})$ to rise together. In this regime, $P(\text{``No''})$ can exceed the absolute threshold~$0.4$ not because a semantic change has occurred, but because the model is in a high-uncertainty state where $P(\text{``Yes''})$ is similarly elevated. The relative formulation self-cancels this common-mode uncertainty: it fires only when the model is \emph{distinctly} more confident in ``No'' than in ``Yes'', naturally suppressing false-positive keyframe insertions under uncertainty. The same property extends to out-of-distribution frames (e.g., near-black or severely blurred frames), where diffuse model confidence causes $S_\mathrm{rel}\approx 0$ by construction, producing a conservative ``no new keyframe'' default without any explicit out-of-distribution detection mechanism.

\section{Channel Robustness Analysis}
\label{appendix:channel}

All experiments in Section~\ref{sec:experiment} are conducted over AWGN channels with SNR~$\in\{6,8,10\}$~dB. This appendix explains why LGVSC's three proposed contributions remain effective under more realistic channel conditions, including Rayleigh fading.

\textbf{Architectural decoupling of semantic and channel layers.}
LGVSC maintains a strict separation between the semantic/source coding layer and the physical channel layer:
\begin{itemize}
  \item The \emph{semantic layer} (PSSS, SKEM, DSA) operates on extracted features and reconstructed video frames. It interacts with the channel layer only through the received SNR delivered to the decoder.
  \item The \emph{channel layer} consists of two independent modules: NTSCC~\cite{NTSCC} for keyframe transmission and LDPC with standard modulation for text descriptions and side information.
\end{itemize}
The three proposed contributions are therefore \emph{orthogonal to the channel model}: their performance is fully determined by the received SNR at the decoder input, regardless of whether that SNR arises from AWGN, Rayleigh fading, or any other propagation model.

\textbf{Channel-layer robustness is established in prior literature.}
NTSCC~\cite{NTSCC} demonstrates robust performance under varying channel conditions, where its learned nonlinear transform naturally adapts to SNR variations through the rate-distortion trade-off. LDPC coding with standard link adaptation (rate matching and modulation order selection) achieves reliable delivery across AWGN and fading channels through well-established coding theory. Under a fading channel, the instantaneous SNR fluctuates around a mean; our experiments at fixed SNR values of $6$, $8$, and $10$~dB directly characterize the semantic quality at the corresponding operating points, effectively covering the relevant quality range of a fading distribution.

\textbf{Future directions.}
End-to-end evaluation under Rayleigh fading with mobility effects, as well as validation on hardware software-defined radio platforms, are important directions for future investigation. The architectural decoupling described above means that such an evaluation would primarily assess the channel-layer modules (NTSCC, LDPC), while leaving the three semantic-layer contributions unchanged.

\section{Ablation Completeness: Analysis of All Module Combinations}
\label{appendix:ablation}

Section~\ref{sec:experiment} compares SKEM+DSA against SKIM+SFA. This appendix explains why the two remaining combinations (SKIM+DSA and SKEM+SFA) are either redundant or architecturally infeasible, making the two evaluated settings the only meaningful experimental configurations.

Table~\ref{tab:ablation_all} summarizes all four possible combinations of \{SKIM, SKEM\}~$\times$~\{SFA, DSA\}.

\begin{table}[htbp]
\caption{Analysis of all four module combinations.
\label{tab:ablation_all}}
\begin{center}
\scriptsize
\renewcommand{\arraystretch}{1.3}
\begin{tabular}{|>{\centering\arraybackslash}m{1.7cm}
                |>{\centering\arraybackslash}m{1.8cm}
                |>{\centering\arraybackslash}m{1.5cm}
                |>{\centering\arraybackslash}m{2.6cm}|}
\hline
\textbf{Combination} & \textbf{Status} & \textbf{Performance} & \textbf{Reason} \\
\hline
SKIM + SFA & Evaluated (baseline) & Lower bound & Fixed-interval extraction; fixed-length adapter \\
\hline
SKIM + DSA & Equiv.\ to SKIM+SFA & $\equiv$ SKIM+SFA & SKIM produces equal-length segments; DSA reduces to SFA \\
\hline
SKEM + SFA & Architecturally incompatible & N/A & Variable-length SKEM output cannot match fixed-length SFA input \\
\hline
SKEM + DSA & Evaluated (proposed) & Upper bound & Semantic-guided extraction; variable-length adapter \\
\hline
\end{tabular}
\end{center}
\end{table}

\textbf{SKIM+DSA $\equiv$ SKIM+SFA.}
SKIM divides a video into a preset number of equal-length segments; because the segment count is fixed in advance, all segments have identical length by construction. DSA dynamically adjusts the VAE latent dimension based on segment length; however, when every segment has the same fixed length, DSA loads the same latent dimension at every step, which is functionally identical to SFA. The neural network's input/output structure and all parameter values are therefore the same in both cases, so SKIM+DSA and SKIM+SFA produce identical metrics within numerical precision.

\textbf{SKEM+SFA is architecturally incompatible.}
SKEM produces segments of variable length because different semantic events span different numbers of frames. SFA maintains fixed latent-space dimensions across all segments. Combining SKEM (variable-length output) with SFA (fixed-length input) would require either (a)~truncating longer segments, which loses semantic content, or (b)~padding shorter segments with zeros or repeated frames, which introduces reconstruction artifacts. Neither option constitutes a meaningful or fair comparison. This incompatibility is precisely the motivation for designing DSA: DSA exists because SKEM requires a variable-length-aware adapter that SFA cannot provide.

\textbf{Attribution of performance gains.}
Since SKIM+DSA $\equiv$ SKIM+SFA, replacing SKIM with SKEM while keeping the adapter accounts for the entire performance gain of SKEM+DSA over SKIM+SFA. This gain is therefore fully attributable to SKEM's semantic-aware keyframe selection. DSA's contribution is architectural: it is the enabler that allows SKEM's variable-length outputs to be processed by the world model, unlocking arbitrary-length video transmission. The SKEM+SFA incompatibility itself is the cleanest proof of DSA's architectural necessity.

\end{document}